\documentclass[aps,prd,superscriptaddress,12pt,showpacs,notitlepage]{revtex4}
    \usepackage{graphicx}
    \usepackage{amsmath,amssymb,mathrsfs}
\usepackage{epsf}
\usepackage{epsfig}

\newcounter{fig}   \newcommand{\lbfig}[1]{\refstepcounter{fig}
\label{#1} }

\newcommand{\Tr}{{\rm Tr}}

\newcommand{\bea}{\begin{eqnarray}}
\newcommand{\eea}{\end{eqnarray}}
\newcommand{\be}{\begin{equation}}
\newcommand{\ee}{\end{equation}}

\newcommand{\re}[1]{(\ref{#1})}

\newcommand{\bb}{\bibitem}

\newcommand{\eqn}{\begin{eqnarray}}
\newcommand{\eqnx}{\end{eqnarray}}

\begin{document}

\title{Hairy black holes in the general Skyrme model}

\author{C. Adam}
\affiliation{Departamento de F\'isica de Part\'iculas, Universidad de Santiago de Compostela and Instituto Galego de F\'isica de Altas Enerxias (IGFAE) E-15782 Santiago de Compostela, Spain}
\author{O. Kichakova}
\affiliation{Institut f\"ur Physik, Universit\"at Oldenburg, Postfach 2503 D-26111
Oldenburg, Germany}
\affiliation{Department of Mathematics and Statistics, University of Massachusetts,
Amherst, Massachusetts 01003-4515, USA}
\author{Ya. Shnir}
\affiliation{Department of Theoretical Physics and Astrophysics, BSU, Minsk, Belarus}
\affiliation{BLTP, JINR, Dubna, Russia}
\author{A. Wereszczynski}
\affiliation{Institute of Physics,  Jagiellonian University,
Lojasiewicza 11, Krak\'{o}w, Poland}

\begin{abstract}
We study the existence of hairy black holes in the generalized Einstein-Skyrme model. It is proven that in the
BPS model limit there are no hairy black hole solutions, although the model admits gravitating (and flat space) solitons.
Furthermore, we find strong evidence that a necessary condition for the existence of black holes with Skyrmionic
hair is the inclusion of the Skyrme term $\mathcal{L}_4$. As an example, we show that there are no hairy black holes
in the $\mathcal{L}_2+\mathcal{L}_6+\mathcal{L}_0$ model and  present a new kind
of black hole solutions with compact Skyrmion hair in the $\mathcal{L}_4+\mathcal{L}_6+\mathcal{L}_0$ model.
\end{abstract}
\maketitle

\section{Introduction}
The Skyrme model \cite{skyrme} is a candidate for the low-energy effective theory of Quantum Chromodynamics in the non-perturbative regime.
In its simplest version, it contains only pionic degrees of freedom $\vec{\pi}$ encoded into an $SU(2)$ valued matrix
field $U$. On the other hand, all other color singlet states - baryons and atomic nuclei - appear as coherent excitations
of the pionic field, i.e., as topological solitons, called Skyrmions \cite{skyrmions}. This is possible due to an identification between
the topological index of Skyrmions and the baryon charge. From the baryonic and nuclear matter physics point of view, the
attractiveness of the Skyrme proposal originates in the fact that all properties of particles, nuclei and nuclear matter
can be derived from the microscopic (pionic) effective action.

The most general Skyrme model which obeys the natural requirements of Poincare invariance and existence of the usual
Hamiltonian consists of four terms

\begin{equation}
\label{lag}
\mathcal{L} \equiv \mathcal{L}_2+\mathcal{L}_4+\mathcal{L}_6+\mathcal{L}_0,
\end{equation}
where $\mathcal{L}_0$ is a potential term, and (here, $a$ and $b$ are non-negative coupling constants; further, we use the 
$(-,+,+,+)$ metric sign convention)
\be
\mathcal{L}_2 = a \mbox{Tr\;} L_\mu L^\mu, \;\; \mathcal{L}_4=b \mbox{Tr\;} [L_\mu, L_\nu]^2, \;\;
L_\mu = U^\dagger \partial_\mu U
\ee
are the sigma model and Skyrme parts, respectively.
Here $L_\mu = U^\dagger \partial_\mu U$ is the $su(2)$-valued left current, associated with
the $SU(2)$-valued  field $U = \sigma \cdot {\mathbb I} + i \pi^a \cdot \tau^a$.
The quartet of fields $(\sigma, \pi^a)$ is restricted to the surface
of the unit sphere, $\sigma^2+ \pi^a \cdot \pi^a =1$.

Further, there is a sextic term ($\lambda$ is a coupling constant)
\be
\label{sextic}
\mathcal{L}_6=\pi^4 \lambda^2 \mathcal{B}_\mu \mathcal{B}^\mu
\ee
where
\begin{equation}
\label{top-curr}
\mathcal{B}^\mu  = \frac{1}{24 \pi^2 \sqrt{-g}}\varepsilon^{\mu\nu\rho\sigma} \Tr (L_\nu L_\rho L_\sigma) \, , \;\;\;
B=\int d^3 x \sqrt{-g} \mathcal{B}^0
\end{equation}
is the topological current and $B$ is the baryon charge. These four terms are not only mathematically distinct but
represent different phenomena and are responsible for different features of nuclear matter. The first two terms,
$\mathcal{L}_2$ and $\mathcal{L}_4$, describe the kinetic part and the two-body interaction, respectively, and take into
account a perturbative (chiral) limit of the Skyrme model. Skyrmions in such a reduced $\mathcal{L}_2+\mathcal{L}_4$
model are fullerene like objects, i.e., shells, with some discrete symmetries and with a radius scaling like $R \sim \sqrt{B}$.
Therefore, one can say that these terms mainly contribute to the {\it surface} properties of nuclei. In this picture, discrete
symmetries of solitons, which are essential for the computation of the vibrational excitations, emerge due to this surface
part of the model. The sextic term is simply the topological current squared and therefore, being topological in its
nature, seems to be particularly important for phenomena related to coherent excitations. Effectively, it correspond to an inclusion of the
$\omega$ meson (repulsive) interaction \cite{omega} and gives the main contribution to the equation of state at high
pressure or density \cite{eos}. Furthermore, $\mathcal{L}_6$ (together with $\mathcal{L}_0$) is just a field theoretical
realization of a perfect fluid in the Eulerian formulation \cite{term}. This is obviously of high importance for
nuclear matter which, to a large extent, is a perfect fluid. Finally, the potential term takes into account some
non-perturbative mesonic physics, as it breaks the chiral symmetry. Interestingly, the submodel
$\mathcal{L}_6+\mathcal{L}_0$, known as the BPS Skyrme model \cite{BPS} (see also \cite{BPS-other,BPS-rec};
for the inclusion of the sextic term in the full model see \cite{GN}),
provides a physically well-motivated idealization of nuclear matter, which is a perfect fluid with {\it SDiff}
symmetry and zero binding energies (in the classical limit) and clearly contributes to the {\it bulk} properties
of nuclei.

Of course, in the full model, the surface and bulk parts couple to each other and no such clear distinction
is possible. However, one can still show that small binding energies appear due to the BPS part
(this can be achieved by inclusion of the full BPS submodel or by an appropriate choice of the potential \cite{BPS-pot}),
while discrete symmetries emerge from the perturbative part.

As the role played by each term in nuclear physics is reasonably well understood, one can ask the same question
in an astrophysical context. This program has been recently started, leading to a beyond mean-field description
of neutron stars within the BPS Skyrme theory. As the BPS model provides the leading contribution to many observables
at high density, it is reasonable to approximate the liquid core of neutron stars by this part of the full Skyrme action.
Such a reduction of the full theory to its BPS limit enables to perform exact computation of properties of neutron
stars, without any mean-field approximation \cite{BPS-stars}. At lower densities, the other terms of the full theory
start to play a role, and more complicated crystal-like structures emerge \cite{sk-stars}.

Here, we investigate black hole solutions in the general Skyrme model. Obviously, as the Skyrme model is
a model of nuclear matter and, after coupling to gravity, a model of neutron stars, one can in principle describe
a merging of such Skyrmionic stars and a collapse to a black hole solution. During such a collapse, some matter
can escape, forming Skyrmionc hair.  As the full dynamics of this process is beyond the scope of the present work,
we want to study the relation between the existence of hairy black holes (representing a finite amount of matter
which did not collapse to a black hole) and the presence of specific terms in the Skyrme model. In other
words, we search for a specific term in the full action which may be responsible for the emergence of Skyrmionic hair.
Let us notice that the (non)existence of hairy black holes is an interesting theoretical problem of its own.
In the context of Skyrmions, such solutions have been extensively studied for the massless
$\mathcal{L}_2+\mathcal{L}_4$ submodel \cite{Luckock:1986tr,Droz:1991cx,bizon}. However, very recently it has been observed by
Gudnason et. al. \cite{bjarke} that, for a specific potential, there are no hairy black holes in the BPS Skyrme model.
This motivated us to study more generally under which conditions hairy black holes do exist in Skyrme type theories.

The dynamics of the system is given by the following action
\be
\label{action}
S=\int d^4 x \sqrt{-g} \left(\frac{R}{16\pi G} + \mathcal{L}\right)
\ee
Here, the gravity part is the usual Einstein--Hilbert
action with curvature scalar $R$, $g$ denotes the determinant
of the (dynamical) metric $g_{\mu \nu}$ and $G$ is the gravitational constant.
Thereafter we mainly restrict 
to the $B=1$ sector and assume the hedgehog (spherically symmetric) ansatz for the matter field
\be
\label{hedgehog}
U=\cos f +i\sin f \; \vec{n}\cdot \vec{\tau}
\ee
where $\vec{\tau}$ are Pauli matrices, $f(r)$ is the profile function of the Skyrmion and
\be
\vec{n}=(\sin \theta \cos \phi, \sin \theta  \sin \phi, \cos \theta) .
\ee

For the metric is in Schwarzschild form, we choose the same parametrization as in \cite{bjarke},
\be
\label{metric-1}
ds^2=-A(r)e^{2\delta (r)} dt^2+\frac{1}{A(r)} dr^2 + r^2(d\theta^2+\sin^2 \theta d\phi^2)
\ee
where the metric function $A$ is assumed to possess a horizon at $r_H$
\be
A(r)=\left( 1-\frac{r_H}{r}\right) \;\;\ \mbox{for} \;\;\; r \rightarrow r_H.
\ee
\section{Black holes in the BPS Skyrme model}
\noindent Let us now consider the BPS part of the full model i.e., the BPS Skyrme model \cite{BPS}
(for recent results concerning the model see \cite{BPS-rec})

\begin{equation}
\mathcal{L}_{BPS} \equiv \mathcal{L}_6+\mathcal{L}_0 \equiv
\pi^4 \lambda^2 \mathcal{B}_\mu \mathcal{B}^\mu- \mu^2 \mathcal{U},
\end{equation}
Since there are solitons for any value of the topological charge in both no-gravity and gravity cases,
one could expect that there should also exist black holes with such BPS Skyrmionic hair. However, as
we prove below, this does not happen. In fact, this feature has been very recently observed in the
BPS Skyrme model with {\it a very particular} potential by Gudnason et. al. \cite{bjarke}. Here we prove
that this is a general property of the BPS Skyrme model for any single vacuum potential $\mathcal{U}$.

To show this surprising result, it is enough to analyze the field equation for the profile function of a Skyrmion
of charge $B$
\be
 \frac{B^2\lambda^2}{2 r^2 } \sin^2 f \partial_r \left(\frac{1}{r^2} e^\delta A \sin^2 f f_r  \right)
  - e^\delta \mu^2 \mathcal{U}_f=0,
\ee
where we extended the hedgehog ansatz to an axially symmetric generalization
\be
\vec{n}=(\sin \theta \cos B\phi, \sin \theta  \sin B\phi, \cos \theta)
\ee
which {\it is consistent} with the full Einstein-BPS Skyrme equations of motion.
Then,
\be
 \partial_r \left(\frac{1}{r^2} e^\delta A \sin^2 f \; f_r  \right) =
 \frac{2  \mu^2}{B^2\lambda^2} r^2 e^\delta \frac{\mathcal{U}_f}{\sin^2 f},
\ee
which after multiplication by $f$ can be written as
\be
 \partial_r \left(\frac{1}{r^2} e^\delta A f \sin^2 f \; f_r  \right)  - \frac{1}{r^2} e^\delta A \sin^2 f \;
 f_r^2  =  \frac{2  \mu^2}{B^2\lambda^2} r^2 e^\delta \frac{f \mathcal{U}_f}{\sin^2 f}.
\ee
Now we integrate this equation from the horizon $r_h$ to $R$, where the Skyrmion field reaches its vacuum
value $f(R)=0$ (recall that the distance $R$ can be finite for compactons, however in is infinite for usual solitons)
\be
  \left. \frac{1}{r^2} e^\delta A f \sin^2 f \; f_r  \right|_{r_h}^R=
  \int_{r_h}^R dr  \frac{1}{r^2} e^\delta A \sin^2 f \; f_r^2  + \int_{r_h}^R dr
  \frac{2  \mu^2}{B^2\lambda^2} r^2 e^\delta \frac{f \mathcal{U}_f}{\sin^2 f}.
\ee
At the horizon we have $A(r_h)=0$, $\delta=1$. Then, assuming that the baryon density is a regular function everywhere
we get that at the horizon, $\sin^2 f \; f'(r_h) \neq \infty$, and at the boundary (where $f(R)=0$),
$\sin^2 f \; f'(R) \neq \infty$. This implies that the l.h.s. is 0.  On the other hand, the r.h.s. is positive
under the assumption that $\mathcal{U}_f \geq 0$. In principle one can consider the situation where at the horizon
\be
\sin^2 f f_r \sim \frac{1}{A}
\ee
which would lead to a positive l.h.s.. But it gives a nonintegrable singularity for the energy density (and baryon density)
at the horizon and, therefore, is not physically acceptable.

As far as we know, this is a unique example of a theory which has static topological solitons both with and without
gravity, but does not have hairy black holes \footnote{On the other hand it is known that there are no black hole analogues
of the regular boson star configurations, which are non-topological solitons in the model of
complex scalar fields minimally coupled to gravity \cite{Pena:1997cy}.}. We want to underline that it is not the non-existence of hairy black holes by
itself which is interesting, but the fact that this happens in a solitonic model with topologically nontrivial solitons.

\section{Generalized Einstein-Skyrme model}
Taking into account the result of the last section,
the natural question arises whether the inclusion of the BPS part into the conventional Skyrme model may have some
impact on the hairy black holes. Let us consider
an extended version of the $SU(2)$ Einstein-Skyrme model \re{lag}
in asymptotically flat 3+1 dimensional space. The matter part of the action is
given by the Lagrangian (here, $c\equiv \pi^4 \lambda^2$ is a non-negative coupling constant)
\be
\label{Skyrme}
\mathcal{L} = a g^{\mu\nu}{\Tr}\left( L_\mu L_\nu\right) + b g^{\mu\nu}
g^{\rho\sigma} {\Tr} \left([L_\mu,L_\rho][L_\nu,L_\sigma]\right) + c
g^{\mu\nu} B_\mu B_\nu + m_\pi^2 \Tr
\left(U-1\right) \, .
\ee
The parameters of the model are, thus, the three coupling constants $a, b, c$ and the pion mass $m_\pi$. Further, we shall restrict our numerical calculations to the pion mass potential, for simplicity.
In the limiting case $c =0$, this action
is just the usual Einstein-Skyrme model with the standard potential term \cite{Luckock:1986tr,Droz:1991cx,bizon}.
In the limit of vanishing coupling constants $a$ and $b$,  the BPS
Einstein-Skyrme $\mathcal{L}_0+\mathcal{L}_6$ model \cite{bjarke} is recovered as a submodel of \re{Skyrme}.

Note that introducing the dimensionless radial coordinate $\tilde r =\sqrt{\frac{a}{b}} r $ and the effective
gravitational coupling constant $\alpha^2={4\pi G a }$, the action of the model \re{Skyrme} can be rescaled as
\be
\label{scaling}
a \to 1;~~ b \to 1;~~ c \to \tilde c =\frac{ac}{b^2}; ~~ m_\pi^2 \to \tilde m_\pi^2= \frac{b m_\pi^2}{a^2} \, .
\ee
Thus, the solutions depend only on the parameters $\tilde m_\pi$, $\tilde c$ and $\alpha$. Clearly, increasing the
parameter $b$ effectively decreases the contribution of the sextic term $\mathcal{L}_6$, while, similar to the
usual Skyrme model, the limit $\alpha^2 \to 0$ corresponds to two situations:
(i) the Newton constant $G \to 0$ (flat space limit), or, (ii) $a \to 0$. Thus, the 
 two-branch structure of regular self-gravitating solutions
known for the conventional Skyrme model
 \cite{Luckock:1986tr,Droz:1991cx,bizon}
is expected to persist.

As before, for the fundamental $B=1$ Skyrmion
we make use of the spherically symmetric hedgehog ansatz \re{hedgehog}. In order to directly compare our results to
the usual pattern of black holes with Skyrmion hair, we now slightly change our parametrization of the metric, parametrizing it as in \cite{Luckock:1986tr,Droz:1991cx,bizon},
\be
\label{metric}
ds^2= - \sigma^2(r) N(r) dt^2 + \frac{dr^2}{N(r)} + r^2 (d\theta^2 + \sin^2 \theta d \phi^2) .
\ee
Here, $N(r) = 1-\frac{2m(r)}{r}$, and the mass function $ m(r)$ may be interpreted as the total mass-energy within
the radius $r$. 

Then the field equations of the model are
\be
\label{general-eqs}
\begin{split}
m_r=&
\frac{\alpha^2}{2}
\left[a \left(\frac{1}{2} r^2 N f_r^2+ \sin^2 f \right)
+b \sin^2 f \left( N {f_r}^2 +\frac{\sin^2 f}{2r^2}\right)
+\frac{c N}{2 r^2}f_r^2 \sin^4 f  +
m_\pi^2 r^2 \sin^2 \frac{f}{2} \right]
; \\
\sigma_r=&
\frac{1}{2} \alpha^2 \sigma  f_r^2\left(a r
+\frac{ 2b \sin^2 f}{r}+\frac{c \sin^4f }{r^3}\right);\\
f_{rr}=&\frac{1}{a +\frac{2 b \sin^2 f}{r^2}+\frac{c \sin^4f}{r^4}}
\biggl\{-a\left[\left(\frac{2}{r}+\frac{N_r}{N}+\frac{ \sigma_r}{\sigma}\right)f_r
-\frac{\sin (2 f)}{r^2 N}\right]\\
&-b\left(-\frac{2 \cos f \sin^3 f}{r^4 N}+\frac{2 f_r \sin^2 f }{r^2}\frac{N_r}{N}
+\frac{{f_r}^2\sin (2 f) }{r^2}+\frac{2 f_r\sin^2 f  }{r^2 }
\frac{ \sigma_r}{\sigma}\right)\\
+&m_\pi^2\frac{\sin f}{2 N}
-c\left(-\frac{2f_r \sin^4 f }{r^5}+\frac{f_r\sin^4 f}{r^4}
\frac{N_r}{N}+\frac{2 {f_r}^2 \cos f \sin^3 f }{r^4}+\frac{f_r \sin^4 f  }{r^4}
\frac{ \sigma_r}{\sigma}\right)\biggr\} \, .
\end{split}
\ee

In the limit $r\to \infty$, the function $m(r)$ is approaching the Arnowitt-Deser-Misner (ADM)
mass of the solution, while
the Skyrme field goes to zero, $f\to 0$.
Note that, similar to the case of the usual self-gravitating Skyrmion, the system of equations
 \re{general-eqs} is invariant with respect to the scaling transformation $\sigma \to \lambda \sigma$.
Thus, the function $\sigma$ can be rescaled to approach $\sigma \to 1$ as $r\to \infty$. Furthermore, this function
can be completely eliminated from the equations.

The set of the equations \re{general-eqs} can be constructed numerically when we impose the
appropriate boundary conditions on the event horizon and on spatial infinity.
In the case of regular self-gravitating solitons, the boundary conditions are \footnote{Other self-gravitating multisoliton
solutions can be constructed by imposing $f(0)=2\pi n, ~~ n\in \mathbb{Z}$.  For the sake of compactness we will not discuss these
configurations here.}
\be
\begin{split}
f(\infty) = 0;&\quad N(\infty)=1;\quad \sigma(\infty) = 1; \\
f(0) = \pi;&\quad N(0)=1;\quad \sigma_r(0) = 0.
\end{split}
\ee
More precisely, for the above system of equations (\ref{general-eqs}) there are four independent boundary conditions $f(0)=\pi$, $f(\infty) =0$, $m(0)=0$ and $\sigma (\infty)=1$, corresponding to the four integration constants of the above system. 
The remaining boundary conditions $m_r (0) = m_{rr}(0)=0$, $f_r (0)=0$ and $\sigma_r (0)=0$ are then consequences of the above equations.
We, therefore, expect a discrete set of regular soliton solutions (concretely, we will find zero, one or two solutions) for given, fixed values of the coupling constants.

\section{Gravitating solitons in the generalized Skyrme model}

The solutions are constructed numerically via the usual shooting algorithm, and the
absolute error is lower than $10^{-6}$.  To check our results for correctness, we also
implement the differential equation solver COLSYS based on the Newton-Raphson method \cite{COLSYS}. Here
we employed the compact radial coordinate $x= r/(1+r) \in [0,1]$.

The system of equations \re{general-eqs} admits an extremely rich pattern of
solutions. It depends on four parameters of the extended Skyrme Lagrangian, $a,b,c, m_\pi$,  the
values of the effective coupling constant $\alpha$ and, in the case of the black holes, $r_h$. Rescaling of the overall Lagrangian density \re{scaling} always allows us
to choose $a=b=1$. Without any loss of generality we can take $m_\pi=1$. In the case of regular self-gravitating solutions,
this leaves us with two significant parameters of the model, $\alpha$ and $c$. The choice $c=0$ corresponds to the case of the pure Einstein-Skyrme model.

The dependence of the usual spherically symmetric $B=1$ Skyrmion on gravity has been studied before
\cite{Luckock:1986tr,Droz:1991cx,bizon}.
Let us briefly discuss the properties of the regular self-gravitating Skyrmions in the model \re{Skyrme}.
First, we observe that for all values of the parameters,
a branch of gravitating solutions emerges from the corresponding flat
space configuration, as the effective gravitational coupling $\alpha$ increases from zero. Along this branch,
the metric function $N(r)$ of the gravitating Skyrmion develops a minimum.
For a fixed value of $\alpha$, the minimum of $N(r)$ becomes deeper as the value of parameter $c$
decreases. To clarify the structure of the solution on this branch, we plot in Fig.\ref{f-5} the metric function $N(x)$ and
the  Skyrme field $f(x)$ (remember $x=r/(1+r)$) of the configuration with $a=b=m_\pi=1$ and a few values of $c$ and
gravitational coupling $\alpha=0.15$.
\begin{figure}[hbt]
\lbfig{f-5}
\begin{center}
\includegraphics[height=.32\textheight, angle =-90]{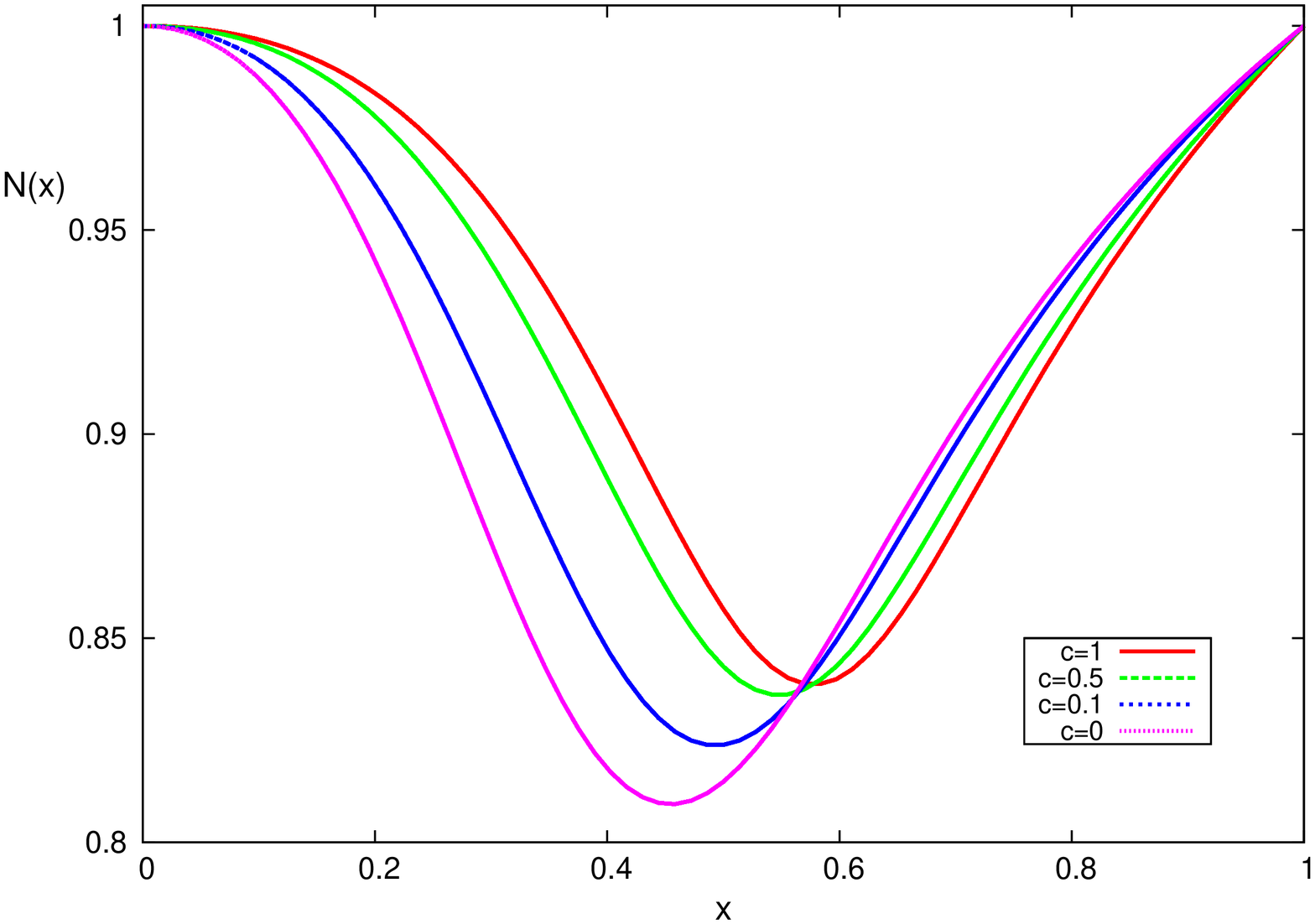}
\includegraphics[height=.32\textheight, angle =-90]{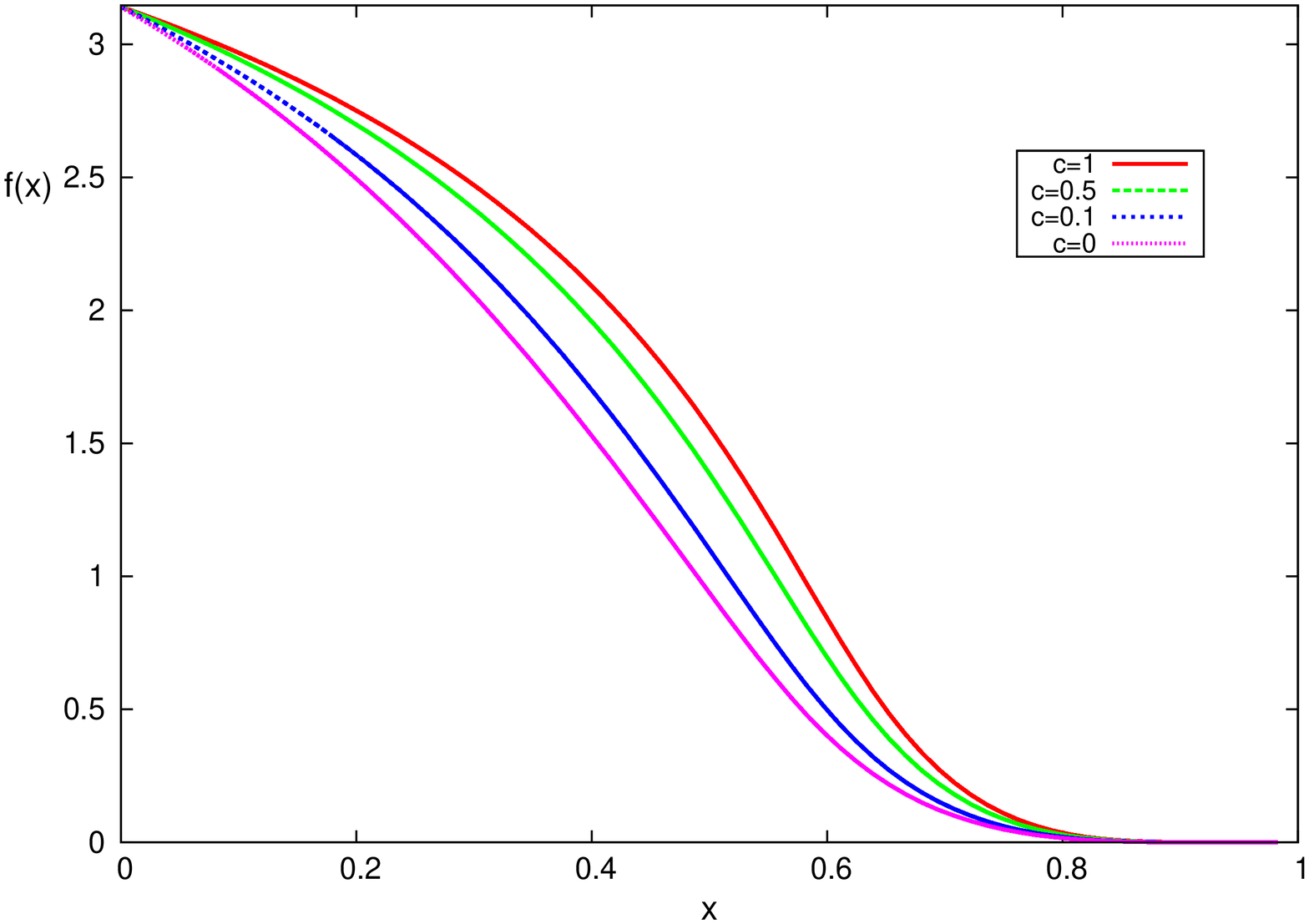}
\end{center}
\caption{\small
The  metric function $N(x)$ of the self-gravitating Skyrmions (left)
and the profile function $f(x)$ (right) are plotted as functions of the compact coordinate $x$
for $a=b=m_\pi=1$ and a few values of $c$ at $\alpha=0.15$.
}
\end{figure}
The minimum of the metric function $N(r)$ along this branch  decreases monotonically  as the coupling constant
$\alpha$ increases. Also, as seen in Fig.~\ref{f-6}, right plot, the dimensionless ADM mass of the configuration, which is given by the asymptotic value of of the
mass function $m(r)$, is increasing. The value of the metric function $\sigma(0)$ decreases, as indicated in Fig.~\ref{f-6}, left plot.

This fundamental (lower mass, and stable) branch extends up to some maximal value of the gravitational coupling $\alpha_{cr}$, beyond which the
gravitational interaction becomes too strong for solutions to
persist. Then this branch bifurcates with a second branch of different type (higher mass, unstable), cf Fig.~\ref{f-6}.
The critical value $\alpha_{cr}$, at which a backbending is observed,
decreases as the value of the parameter $c$ is decreasing. Indeed, the term $\mathcal{L}_6$
effectively corresponds to repulsion in the system, so gravity must be stronger to compensate for it.

\begin{figure}[hbt]
\lbfig{f-6}
\begin{center}
\includegraphics[height=.32\textheight, angle =-90]{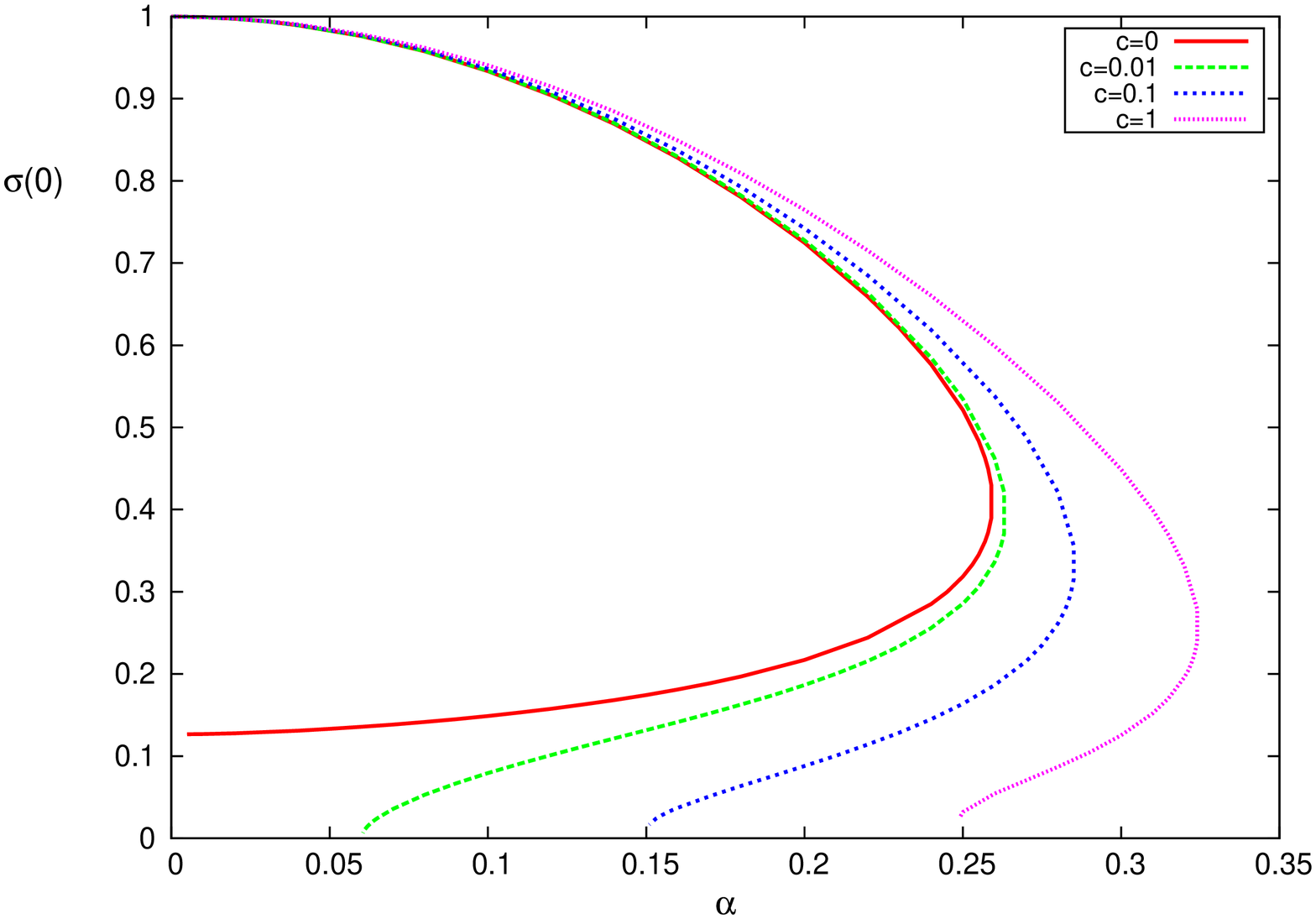}
\includegraphics[height=.32\textheight, angle =-90]{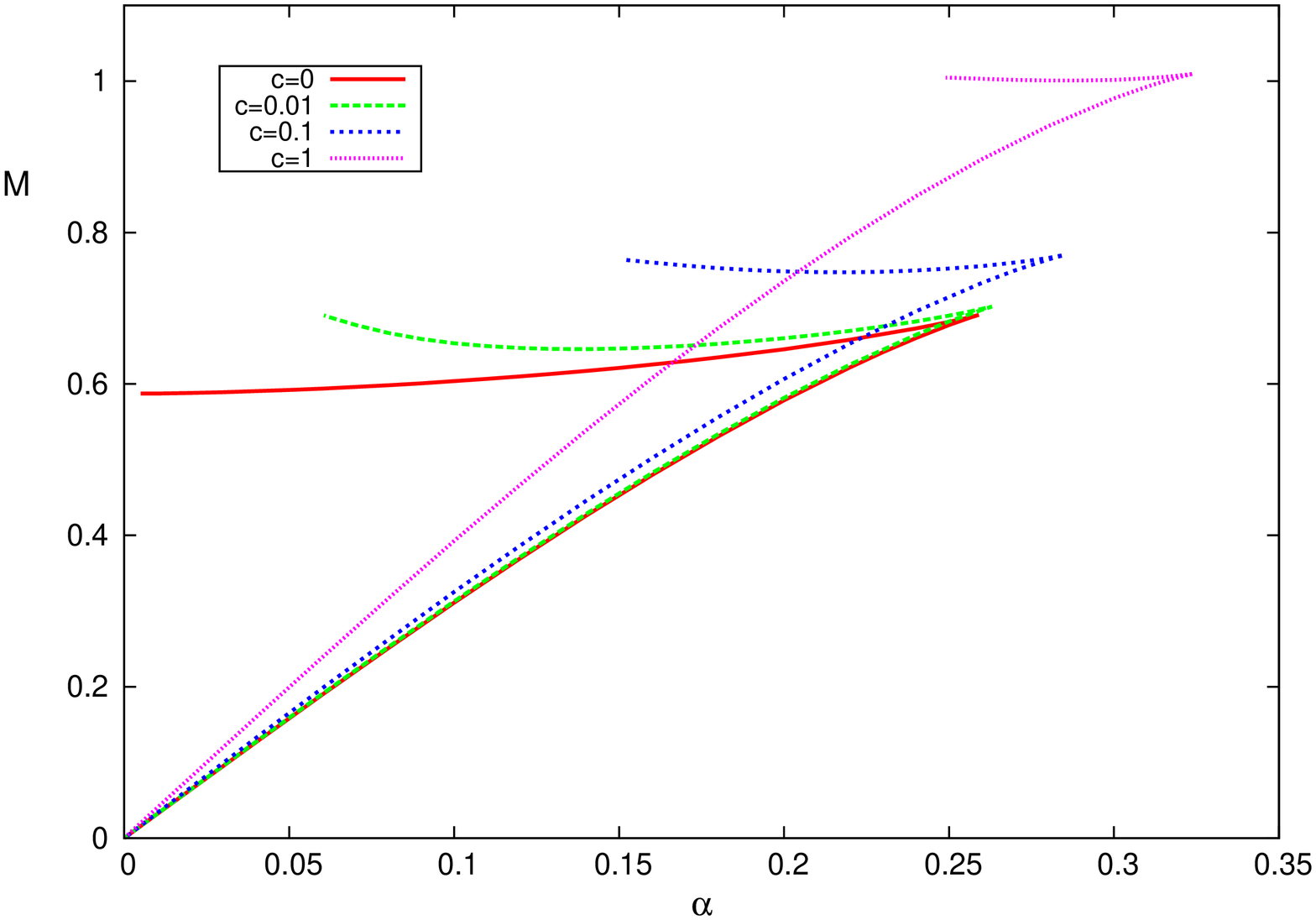}
\end{center}
\caption{\small
The value of  the metric function $\sigma$  of the self-gravitating Skyrmions
at the origin, $\sigma(0)$,  (left) and the ADM masses of the corresponding
solutions (right) are plotted as functions of the coupling constant $\alpha$ for $a=b=m_\pi=1$ and a few values of $c$.
}
\end{figure}

The pattern of evolution along the second (unstable) branch crucially depends on the value of the parameter $c$.
In the limiting case $c=0$, the usual Einstein-Skyrme model is recovered. Then the second
branch extends all the way back to the limiting case $\alpha \to 0$, see Fig.~\ref{f-6}, left plot. Along this second (higher mass) branch,
the solutions become unstable. As is well-known \cite{bizon}, in the limit $\alpha \to 0$ on this branch, the gravitating Skyrmion approaches the
lowest Bartnik-McKinnon (BM) solution of the $SU(2)$ Einstein-Yang-Mills theory \cite{Bartnik:1988am}.

The situation becomes different for $c\neq 0$. In this case, on the second branch, both $\sigma(0)$ and
the minimum of the metric function $N(r)$ decrease further. However, the value of $\sigma(0)$ decreases quicker than
the minimum of $N(r)$, and at some second non-zero critical value of $\alpha$, $\sigma(0) \to 0$, thus the branch terminates at a
singular solution.

This pattern is very similar to the critical behavior observed in the Yang-Mills theory with higher order terms \cite{Brihaye:2002jg}.
Indeed, if the coupling to the sextic term $\mathcal{L}_6$ is non-zero,  the limiting $a\to 0$
solution would correspond to the model with both $\mathcal{L}_4$ and $\mathcal{L}_6$ terms. Then these terms would match the
first two terms in the Yang-Mills hierarchy, which support generalized BM solutions \cite{Brihaye:2002jg}. On the other hand, the existence
of critical singular solutions, yet of a different type, is not unusual in the scalar model with a Skyrme-like term \cite{Radu:2011uj}.

We can also consider a situation where both parameters $a$ and $b$ are decreasing simultaneously, and
the general model is approaching the limiting $\mathcal{L}_0+\mathcal{L}_6$ BPS submodel. The scaling properties of this
submodel are different. If both $a$ and $b$ are set to zero, there is only one parameter $G$ to vary, and
the second, backward branch ceases to exist. Indeed, numerical evaluation shows in this limit the critical value of the
coupling constant $\alpha_{cr}$ increases. Severe numerical difficulties,
which are typical in the case of solitons with compact support, do not allow us to investigate the limiting submodel
using the same numerical scheme. However, there are indications that this branch of gravitating compactons
is not bending backwards.

\section{Hairy black holes in the generalized Skyrme model}

According to the usual arguments (see, e.g. \cite{Volkov:1998cc}), we can expect that there are black hole
generalisations of the regular configurations above, which may exist at least for
small values of the event horizon radius $r_h$.

The surface gravity of static black holes is defined as
\be
\kappa^2= -\frac{1}{4} g^{tt}g^{rr}(\partial_r g_{tt})^2.
\ee
It can be evaluated from the metric functions on the horizon. The temperature of the black hole is proportional to the surface gravity,
$2 \pi T=\kappa$. Thus, for the metric in the usual Schwarzschild
coordinates \re{metric}, we have the Hawking temperature and the entropy of the black hole
\be
\label{temperature}
T = \frac{1}{4\pi} \sigma(r_h)N^\prime(r_h);\qquad S= \frac{A}{4}
\ee
where $A$ is the horizon area, $A = 4\pi r_h^2$.

In the case of black holes,
the boundary conditions on the fields must secure the surface gravity as well as the other curvature invariants
to be finite on the event horizon (the zeroth law of black hole physics).
Thus, we should consider the asymptotic expansion of the fields in the vicinity of $r_h$.
The first terms of the approximate expansion of the metric functions and the profile function of the Skyrmion
in a near-horizon power series expansion in $r-r_h$ read
\be
\label{expansion}
\begin{split}
m \approx & \; \frac{r_h}{2} + m_1 (r-r_h) + \dots \\
\sigma \approx & \; \sigma_h +
\frac{ \alpha^2  \sigma_h \sin^2 f_h J^2 }
{8 r_hHK^2}(r-r_h)
 + \dots \\
f \approx & \; f_h
+\frac{r_h \sin f_h J}{2 HK} (r-r_h) +\dots
\end{split}
\ee
where
\be
\label{JHK}
\begin{split}
m_1 \equiv & \frac{\alpha^2}{4} \left(m_\pi^2 r_h^2 (1-\cos f_h) +
2 a \sin^2 f_h+b\, r_h^{-2}\sin^4 f_h\right) \\
J \equiv & \; m_\pi^2 r_h^4 + \left(b+4 a r_h^2\right) \cos f_h -b \cos (3 f_h) \\
H \equiv & \; a r_h^4 +2 b r_h^2 \sin^2 f_h + c \sin^4 f_h \\
K \equiv & \; 1- 2m_1 .
\end{split}
\ee
Further, $f_h$ and $\sigma_h$ are the values of the Skyrme field and the metric function $\sigma$
at the event horizon. Together with $r_h$, they provide three input parameters in the numerical solution of the system
\re{general-eqs}. On the other hand, we still have to fulfill the two boundary conditions $f(\infty) =0$ and $\sigma(\infty)=1$. We expect, therefore, a one-parameter family of solutions (parametrized, e.g., by the black hole radius $r_h$) for fixed values of the coupling constants. In other words, for a fixed value of $r_h$, we expect a discrete set of hairy black hole solutions (concretely, we shall find zero, one or two solutions, depending on the value of $r_h$).

\subsection{Extremal limit on the unstable branch}

The properties of spherically symmetric black hole solutions linked to Skyrmions
were investigated in \cite{Luckock:1986tr,Droz:1991cx,bizon}, for a review see e.g. \cite{Volkov:1998cc}.
It was shown that in this system
there are hairy black hole solutions, which
can be viewed as bound states of Skyrmions and Schwarzschild black holes \cite{Ashtekar}.
In particular, in the genuine Skyrme model,
for a fixed non-zero value of the gravitational coupling
$\alpha$,
there are always two $r_h$-branches of solutions which bifurcate at a critical, maximal value of $r_h^{\rm cr}$.
Note that the hairy black hole solutions exist in a restricted domain of the $(r_h, \alpha)$ plane.

The branch of solutions which is lower in mass is stable and,
as $r_h \to 0$, this branch merges the corresponding branch of regular self-gravitating Skyrmions,
which has a flat space limit.
The unstable upper (higher mass) branch of the solutions merges the second
$\alpha$-branch of the regular self-gravitating
Skyrmions in the limit $r_h \to 0$, which is linked to the corresponding limiting Bartnik-McKinnon solution as $\alpha \to 0$, thus both branches possess the limit of vanishing event horizon.

The situation is different in the extended Skyrme model  \re{Skyrme}, particularly, for $c\not= 0$, although the bifurcation into two $r_h$ branches continues to hold. Again, for a given $\alpha$,
starting from a regular lower branch solution of the generalized Einstein-Skyrme model \re{Skyrme} at $r_h =0$ and increasing $r_h$, we find a lower $r_h$-branch of black hole solutions.
For $r_h\ll 1$, the solutions resemble small  Schwarzschild black holes placed in the
center of the regular generalized Skyrmion.
This branch extends up to a critical value of  $r_h^{\rm cr}$, at which it bifurcates with the second $r_h$-branch,
as illustrated in Fig.~\ref{f-8}. The Hawking temperature
decreases as $r_h$ increases, while the mass of the solution is increasing with $r_h$. As one can see in Fig~\ref{f-10}, along
this branch both the small
hairy black hole solutions of the extended Einstein-Skyrme model, and the corresponding solution of the truncated
$c=0$ model, behave similarly, the curve $T(r_h)$ is slightly below
the vacuum Schwarzschild solution:  $T_{Schwarz}=\frac{1}{4\pi r_h}$. This branch is stable \cite{Maeda:1993ap}.

\begin{figure}[hbt]
\lbfig{f-8}
\begin{center}
\includegraphics[height=.32\textheight, angle =-90]{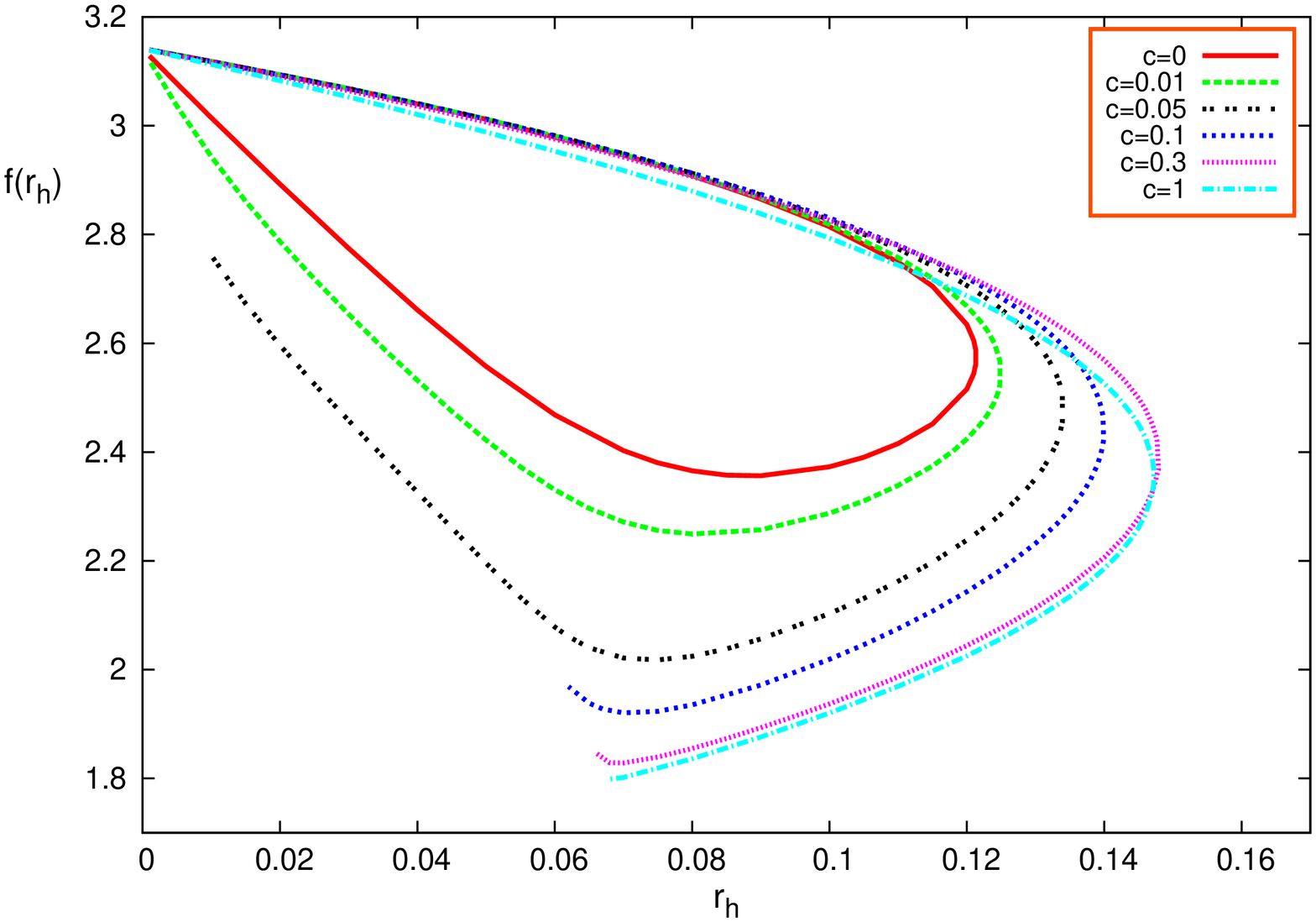}
\includegraphics[height=.32\textheight, angle =-90]{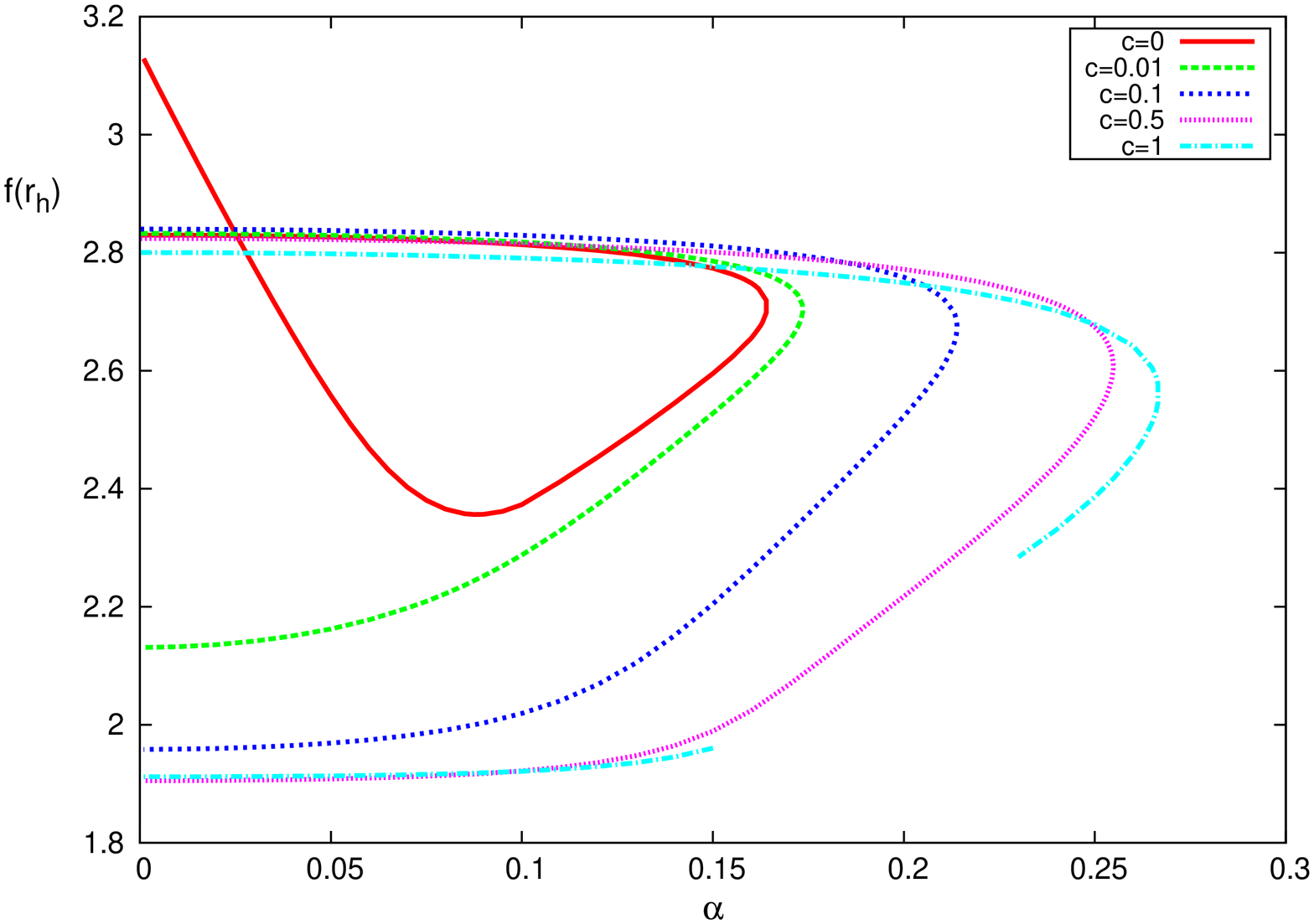}
\end{center}
\caption{\small
\emph{Left panel:}
The value of the Skyrme function $f(r_h)$ at the event horizon
is plotted as a function of the horizon radius $r_h$ for $\alpha=0.05$,  $a=b=m_\pi=1$ and a few values of $c$.
\emph{Right panel:} The value of the Skyrme function $f(r_h)$ at the event horizon
is plotted as a function of the coupling $\alpha$ for  $r_h=0.10$,  $a=b=m_\pi=1$ and a few values of $c$}
\end{figure}

\begin{figure}[hbt]
\lbfig{f-10}
\begin{center}
\includegraphics[height=.32\textheight, angle =-90]{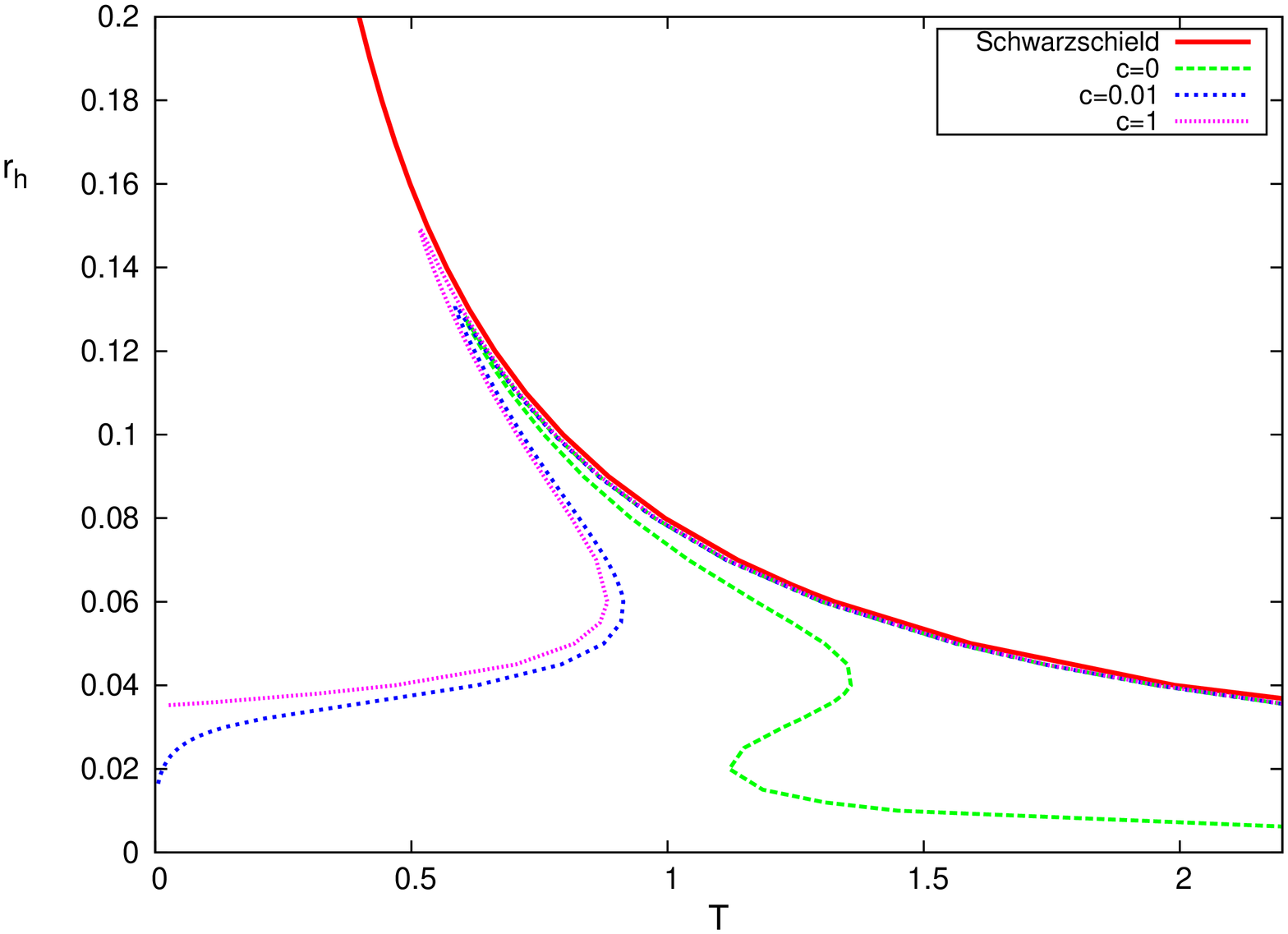}
\includegraphics[height=.32\textheight, angle =-90]{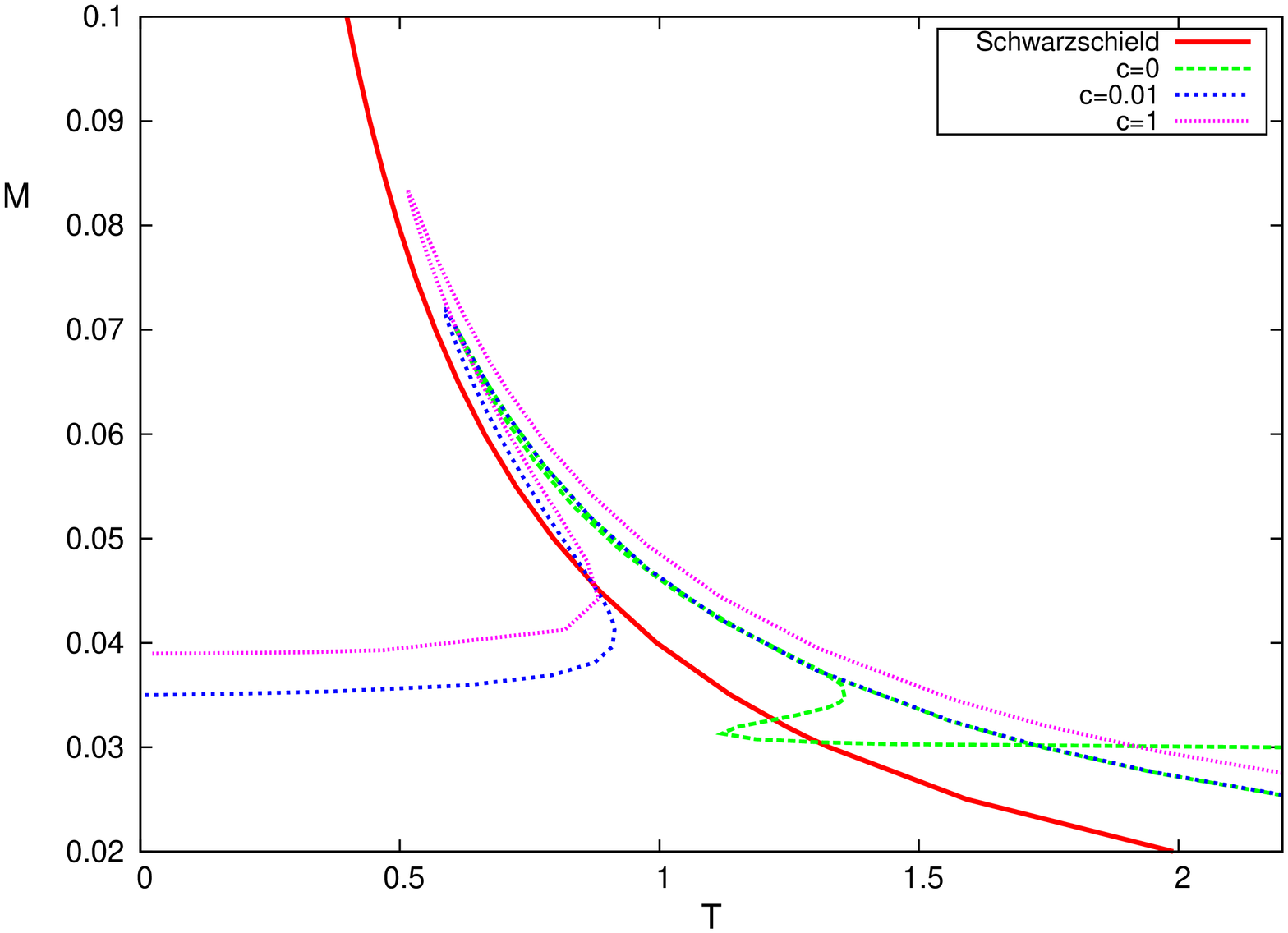}
\end{center}
\caption{\small
\emph{Left panel:}
The Hawking temperature $T$ is plotted as a function of the horizon radius $r_h$
for $\alpha=0.05$,  $a=b=m_\pi=1$ and a few values of $c$.
\emph{Right panel:} The mass of the solutions
is plotted as a function of the temperature for the same set of values of the parameters. The Schwarzschild solution is plotted
as a reference curve.}
\end{figure}

Extending backwards in $r_h$, we find a second branch of solutions, where both branches form a typical cusp at the critical point $r_h^{\rm cr}$.
Along this branch, the patterns of evolution of the hairy black holes in the models with $c=0$ and $c\neq 0$ become different, although
the entropy of all solutions beyond the critical point is higher than on the first (stable) branch.

In the general model \re{Skyrme} (i.e., for $c\not= 0$), the value of the Skyrme function $f(r_h)$ on the horizon continues to decrease monotonically
along the upper branch with decreasing $r_h$, see Figs.~\ref{f-8}. Then it approaches some
finite value at some non-zero value $r_h = r_h^{\rm ex}$, beyond which solutions cease to exist. At this point, the temperature of the hairy black holes approaches zero, as exhibited in
Fig~\ref{f-10}, thus the configuration reaches an extremal, limiting solution.
The formation of this singular (or extremal) point may be better understood by a closer examination of the expansion \re{expansion}.
Indeed, the linear expansion terms both of $\sigma$ and of $f$ in \re{expansion} are divided by the factor $K$ defined in \re{JHK}.
It follows from the definition of $K$ that in the limit of sufficiently small $r_h$, $K$ will approach the value $K=0$ unless $f_h$ approaches the value $f_h \to \pi$ sufficiently quickly, at the same time. On the stable branch (and, for $c=0$, also on the unstable branch) this is precisely what happens, and hairy black hole solutions exist all the way down from $r_h^{\rm cr}$ to $r_h =0$. As explained above, however, for $c\not= 0$, $f_h$ {\em decreases} along the unstable branch with decreasing $r_h$, which seems to imply the formation of the singularity $K=0$. What happens, instead, in the numerical integration, is that for decreasing values of $r_h$ along the upper, unstable branch, the boundary conditions at infinity enforce smaller and smaller values for $\sigma_h$, until the value $\sigma_h =0$ is reached for some nonzero $r_h = r_h^{\rm ex}$, see Fig~\ref{f-sig-bh}. This point corresponds to the limiting (extremal) solution, and no solutions exist along the upper branch for even smaller values of $r_h$.  

\begin{figure}[hbt]
\lbfig{f-sig-bh}
\begin{center}
\includegraphics[height=.42\textheight, angle =-90]{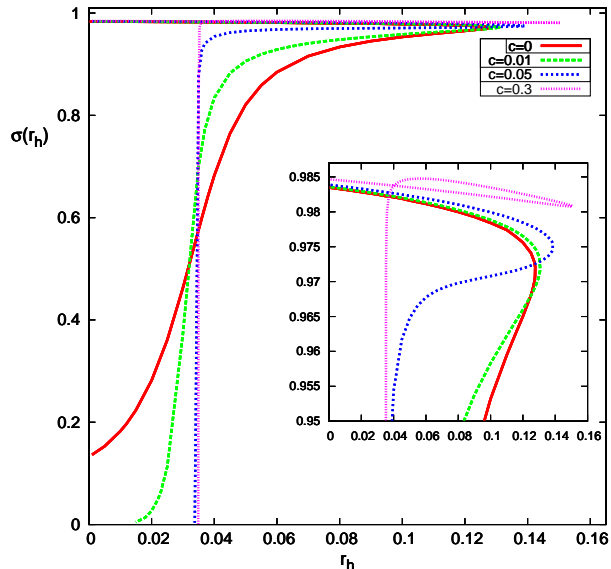}
\end{center}
\caption{\small
The metric function $\sigma$ at the black hole horizon is plotted, both for the stable (lower) and for the unstable (upper) branch,
for $\alpha=0.05$,  $a=b=m_\pi=1$ and a few values of $c$. At the upper branch, the value of $\sigma(r_h)$ goes to zero at some limiting, extremal value $r_h =r_h^{\rm ex}$ if $c\not= 0$, and no solution along the upper branch exists for $r_h < r_h^{\rm ex}$. The jump to zero gets more abrupt for larger $c$.}
\end{figure}

In contrast, for the  Skyrme submodel with $c=0$, the Hawking
temperature does not depend monotonically on $r_h$, see Fig~\ref{f-10}.
For some interval of values of the temperature, there are
three solutions with different masses at a given $T$. Also the limiting behavior of these hairy black hole solutions on the upper branch is different, in the limit $r_h\to 0$ the value of the temperature rapidly grows again, as  shown in Fig~\ref{f-10}.

\subsection{Role of the Skyrme term}

We can further investigate the domain of existence of hairy black holes by varying different coupling constants. First, let us keep the value of the horizon radius $r_h$ fixed,
and vary the coupling $\alpha$. As we observe,
the solutions terminate for some critical values of $\alpha$, see Figs.~3 (right), 9 (left). These critical values of the parameters
$(\alpha, r_h)$ are, again, related to the singularity of the expansion \re{expansion}, as explained above, and the configurations approach the extremal solution
with zero temperature.

To clarify the limiting behavior of solutions, let us now keep all parameters of the model fixed except for the
parameter $b$ which multiplies the Skyrme term $L_4$. Surprisingly, we observe that as $b\to 0$, the critical value
of the
horizon radius $r_h$ is decreasing, and in the
limit where the Skyrme term is absent, we do not find any black hole solutions with Skyrmionic hair.
This pattern is illustrated in Fig.~\ref{f-2}.

\begin{figure}[hbt]
\lbfig{f-2}
\begin{center}
\includegraphics[height=.32\textheight, angle =-90]{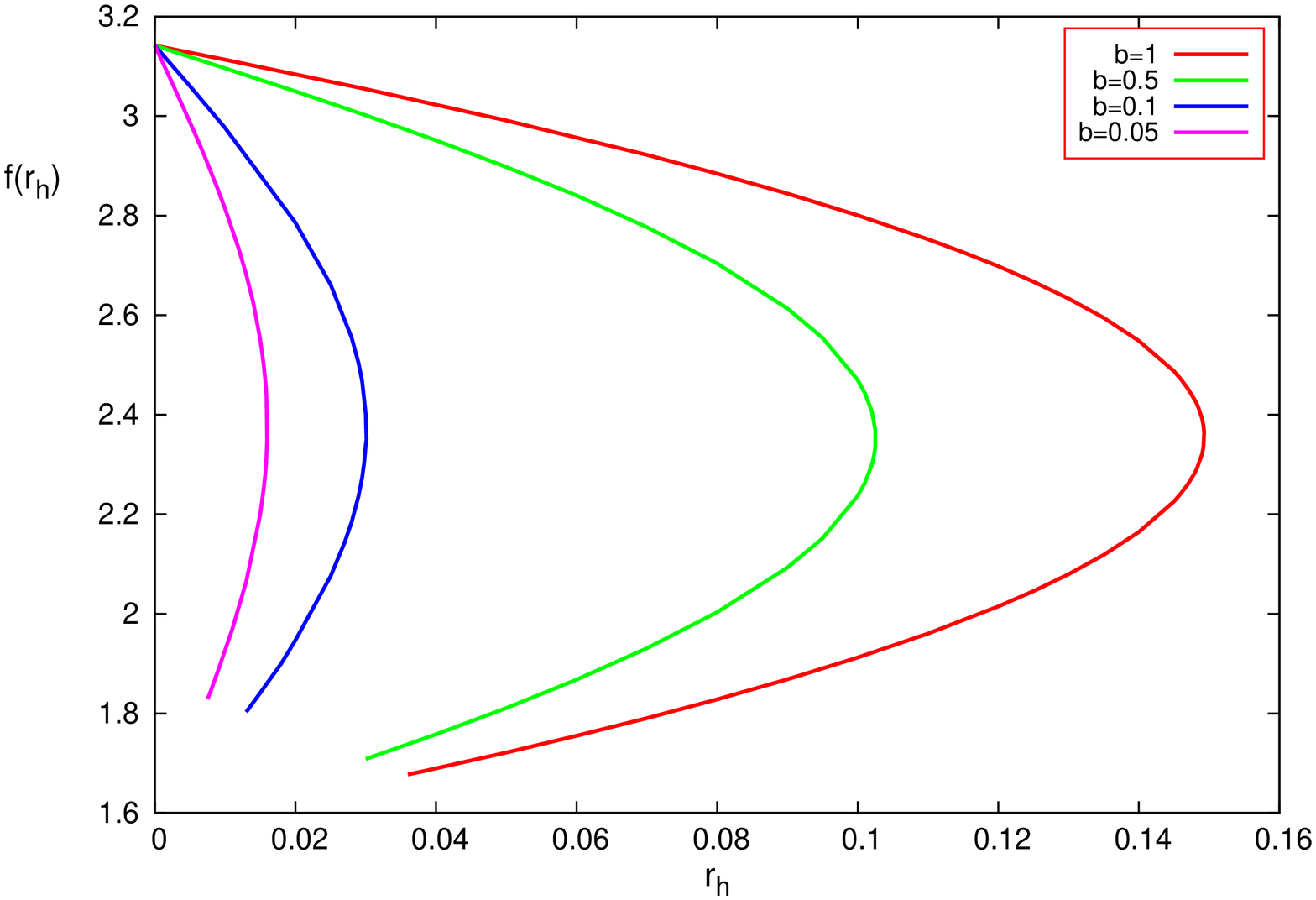}
\includegraphics[height=.32\textheight, angle =-90]{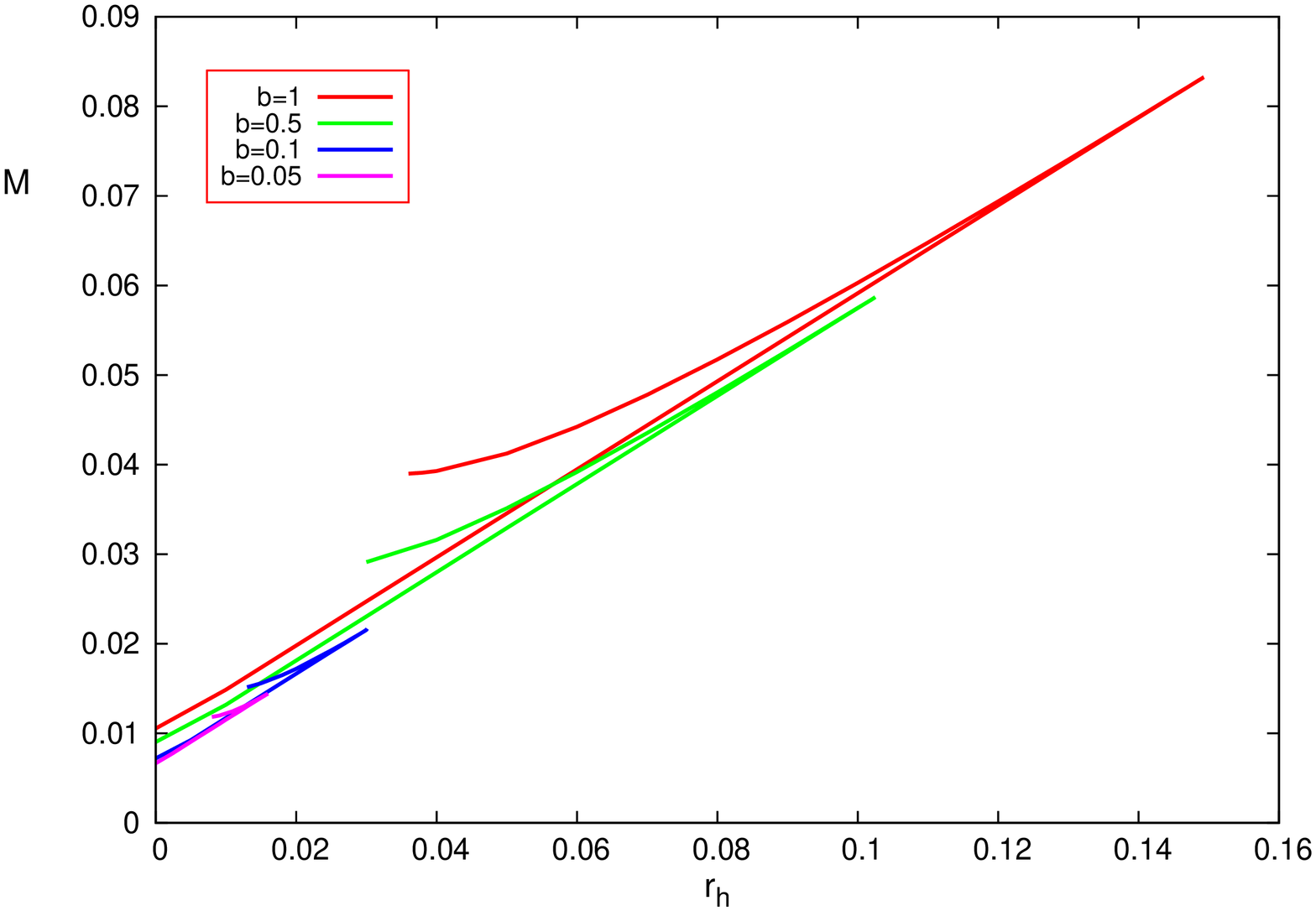}
\end{center}
\caption{\small
The value of the Skyrme function $f(r_h)$ at the event horizon (left) and the ADM mass of the hairy black holes
(right)
are plotted as functions of the horizon radius $r_h$ for $\alpha=0.05$,  $a=c=m_\pi=1$ and a few values of $b$.}
\end{figure}

Thus, analogously to the case of the BPS Skyrme model, the submodel
$\mathcal{L}_0+\mathcal{L}_2+\mathcal{L}_6$ does not support hairy
black holes, although regular self-gravitating topological solitons do exist. Notice that this model contains the
usual kinetic term and, therefore, provides a well defined dynamics, at least for some set of Cauchy data.
One can conclude
that it is not the sigma model part which triggers the hair. On the contrary, it seems that it is the Skyrme term which
is crucial for the existence of hairy black holes.

\begin{figure}[hbt]
\lbfig{f-4}
\begin{center}
\includegraphics[height=.32\textheight, angle =-90]{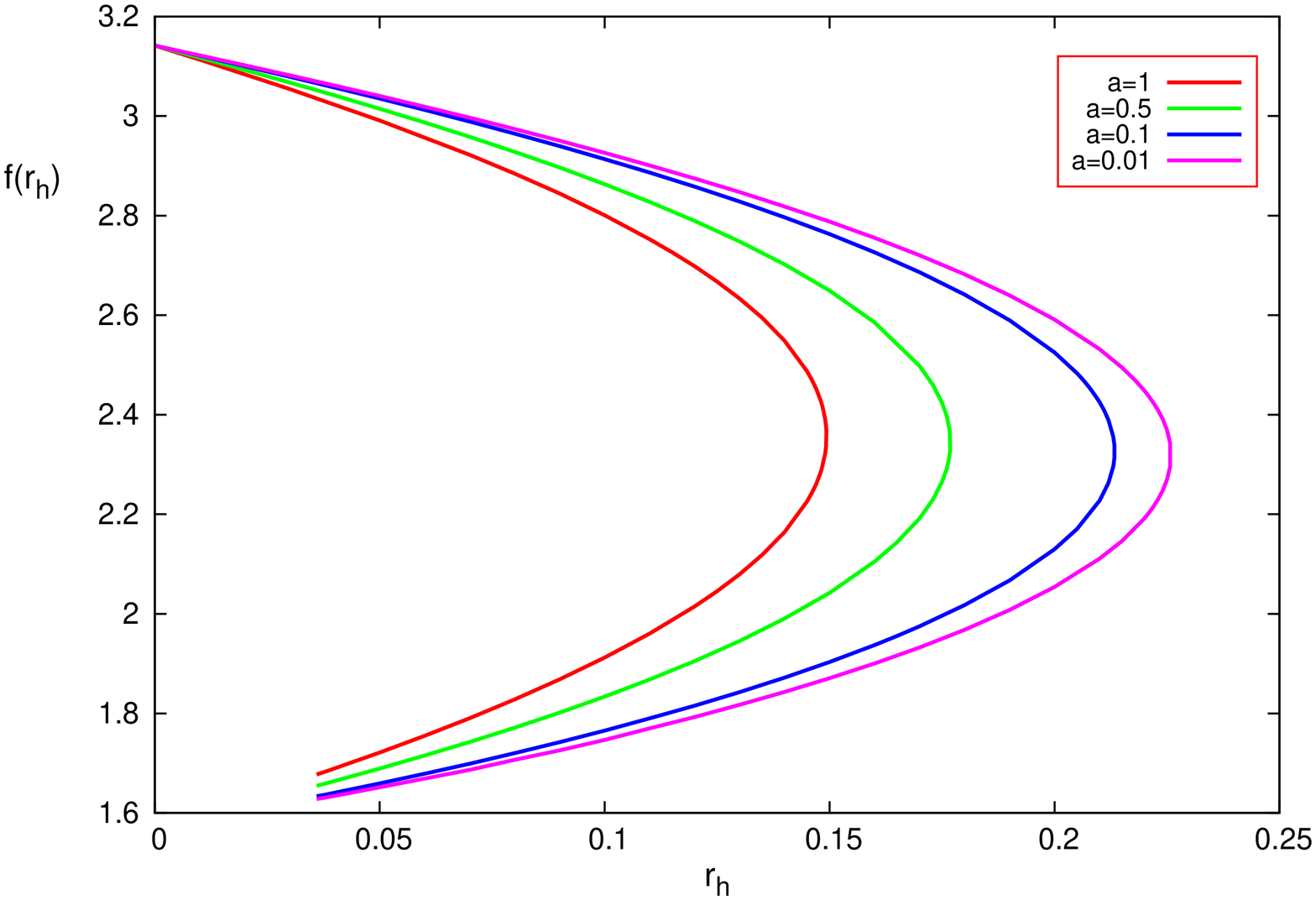}
\includegraphics[height=.32\textheight, angle =-90]{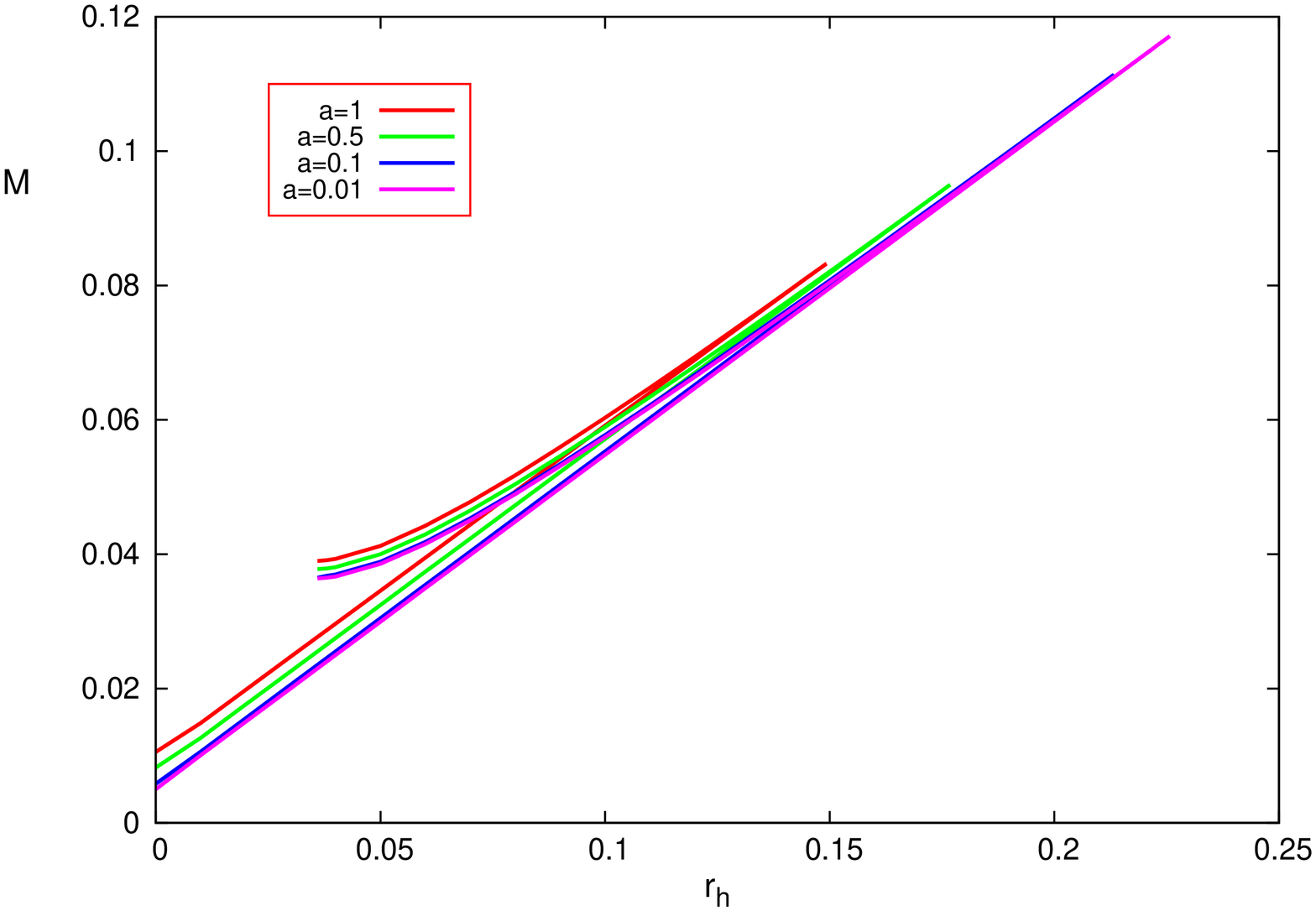}
\end{center}
\caption{\small
The value of the Skyrme function $f(r_h)$ at the event horizon (left) and the ADM mass of the hairy black holes
(right)
are plotted as functions of the horizon radius $r_h$ for $\alpha=0.05$,  $b=c=m_\pi=1$ and some set of decreasing
values of $a$.}
\end{figure}

Let us now consider another limit,
keeping all parameters of the model fixed except for the
parameter $a$ which multiplies the kinetic part $\mathcal{L}_2$.

As seen in Fig.~\ref{f-4}, the critical value of the
horizon radius $r_h$ is increasing as $a$ is decreasing, and the hairy black holes persist in the limit $a\to 0$.
However, such a configuration has \emph{compact hair}. This is a consequence of the fact that
this model, in the flat space-time limit, has compactons as solutions for 
 the usual Skyrme potential assumed here.

In Fig.~\ref{f-1}, right plot, we exhibit the profile function of the Skyrmion on the lower branch
for a few decreasing values of $a$. Clearly, one can see how the configuration approaches the limit of a
hairy black hole with compact support.

Finally, let us consider the situation where both parameters $a$ and $b$ are decreasing simultaneously, and
the general model is approaching the limiting
$\mathcal{L}_0+\mathcal{L}_6$ BPS submodel.
Then the pattern we obtain, becomes qualitatively similar to the one we observed above as $b\to 0$.
In Fig.~\ref{f-9} (right), we represent the results of the analysis of the evolution of the corresponding solutions.

\begin{figure}[hbt]
\lbfig{f-1}
\begin{center}
\includegraphics[height=.32\textheight, angle =-90]{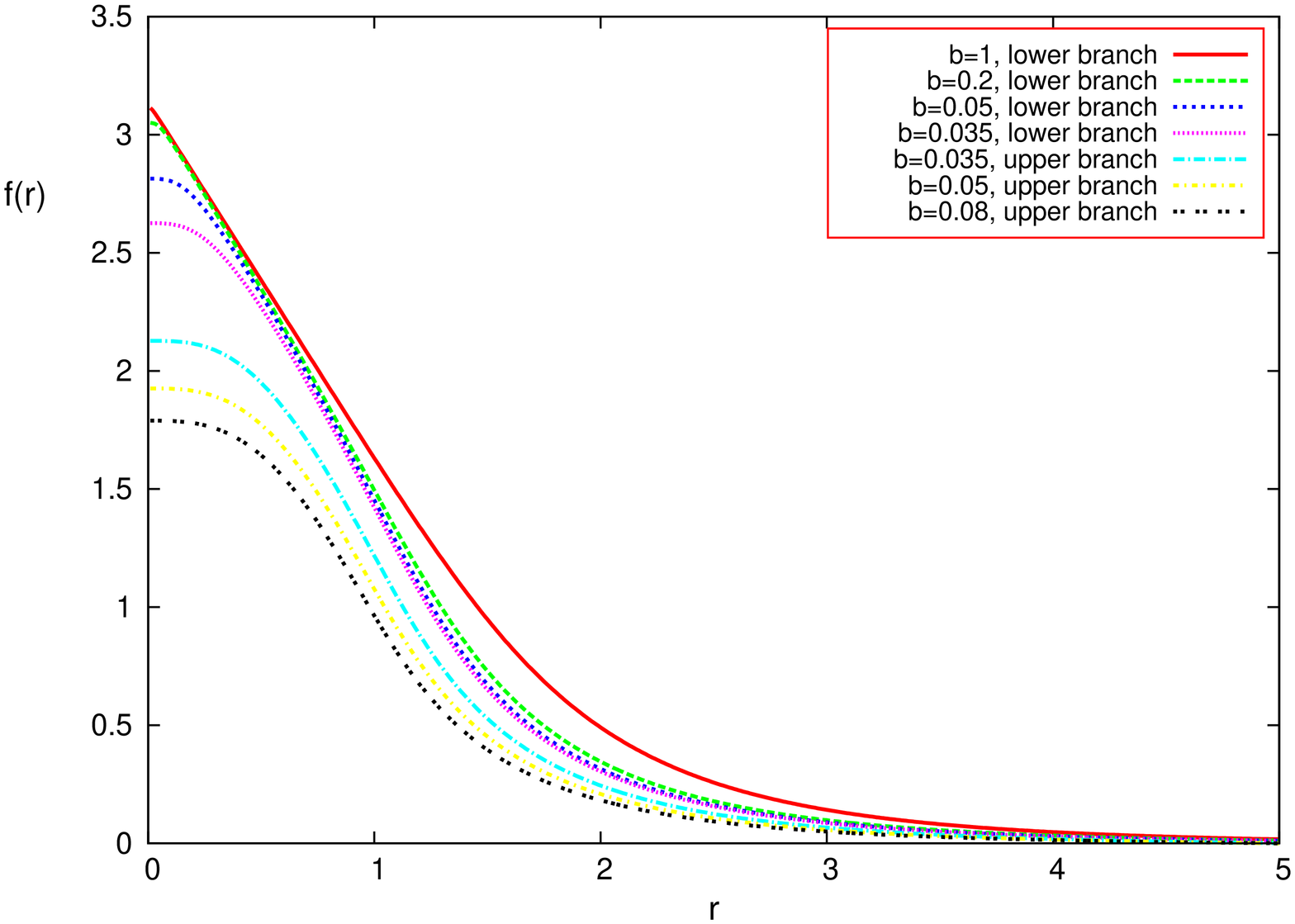}
\includegraphics[height=.32\textheight, angle =-90]{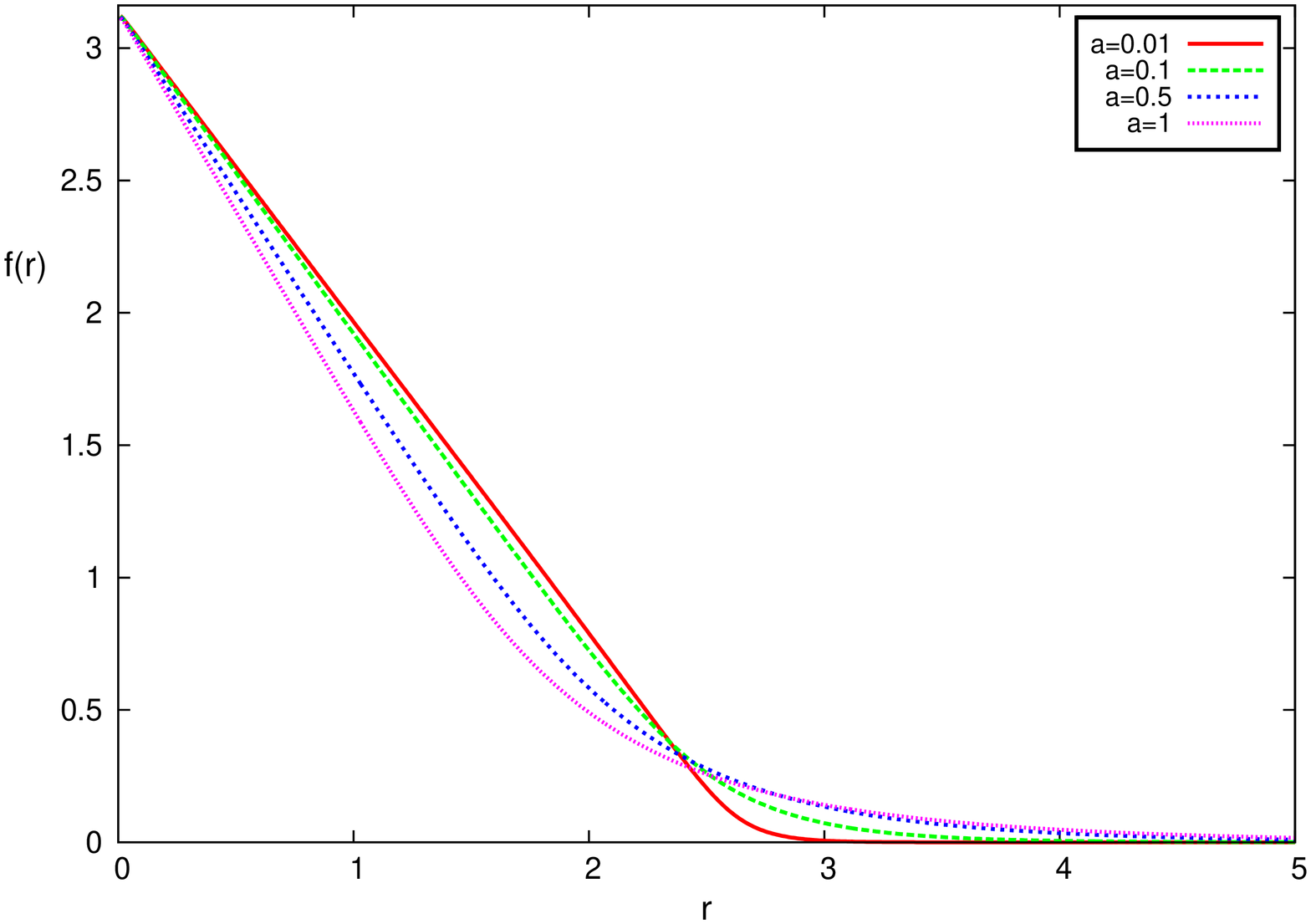}
\end{center}
\caption{\small \emph{Left panel:}
The Skyrme function $f(r)$  is plotted for $\alpha=0.05$ and $r_h=0.01$
for several sets of solutions on the lower and upper branches with  $a=c=m_\pi=1$ and a few values of $b$.
\emph{Right panel:} The Skyrme function $f(r)$  is plotted for $\alpha=0.05$ and $r_h=0.01$
for several sets of solutions on the lower branch with  $b=c=m_\pi=1$ and a few values of $a$.}
\end{figure}

\begin{figure}[hbt]
\lbfig{f-9}
\begin{center}
\includegraphics[height=.32\textheight, angle =-90]{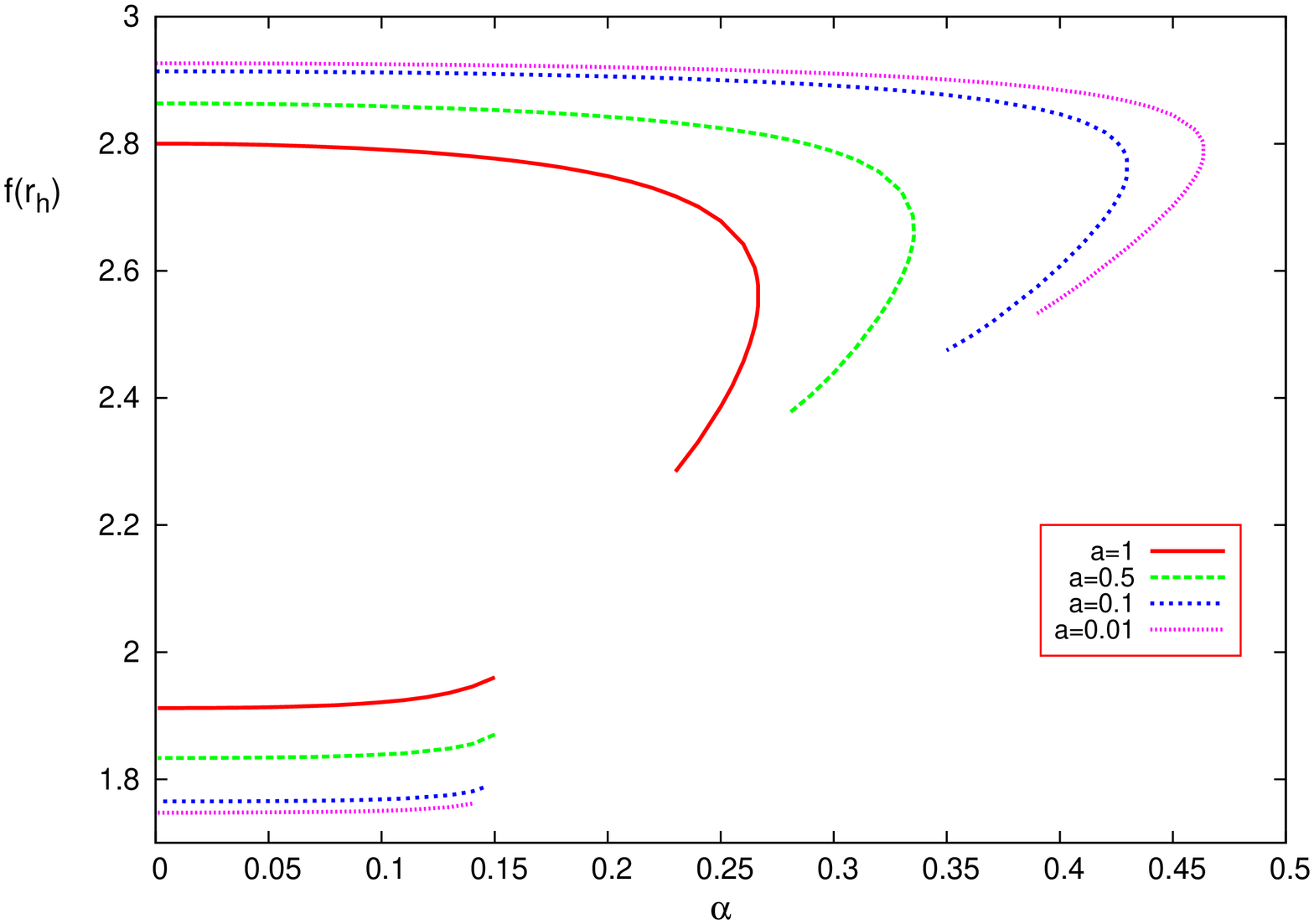}
\includegraphics[height=.32\textheight, angle =-90]{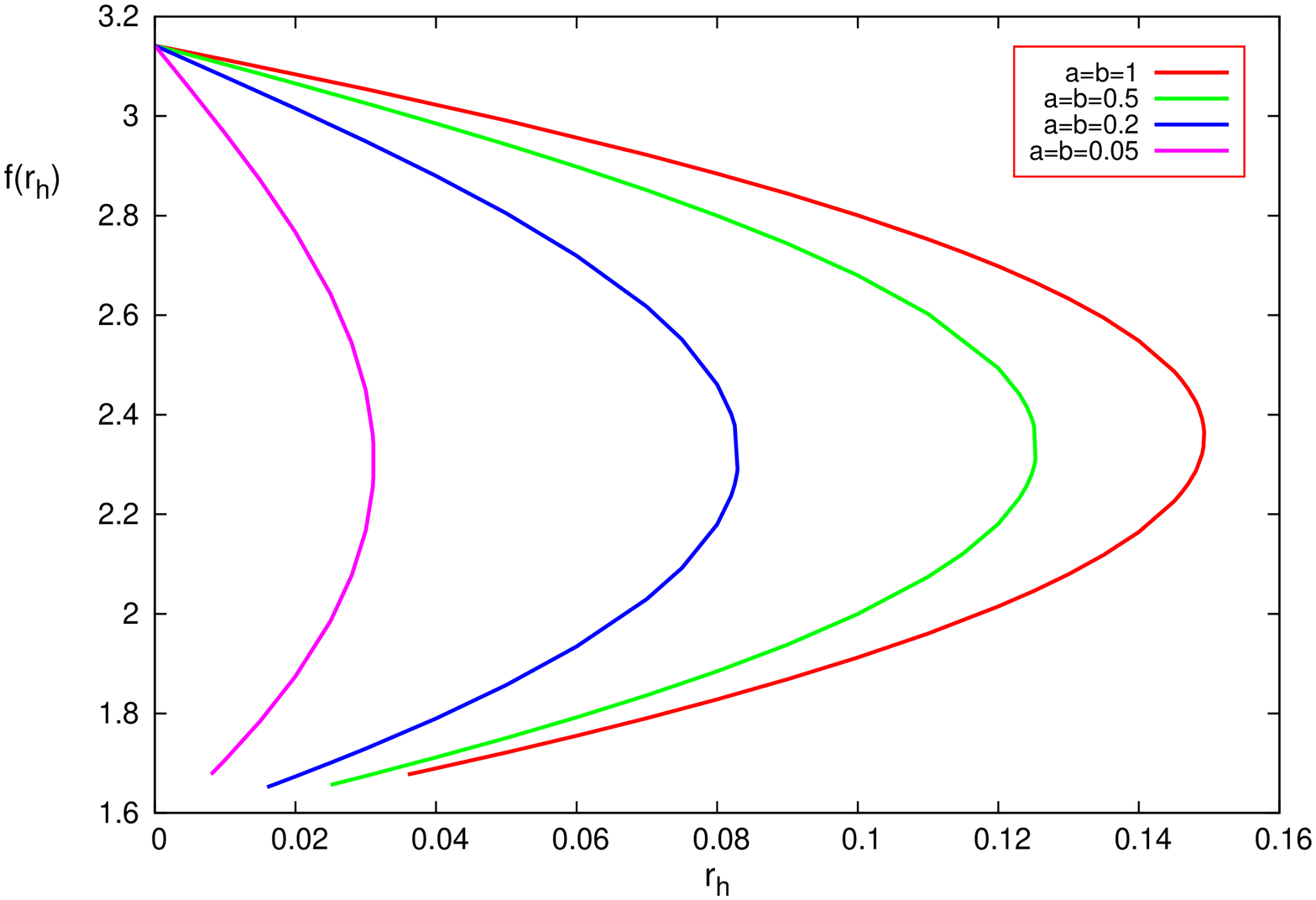}
\end{center}
\caption{\small
\emph{Left panel:} The value of the Skyrme function $f(r_h)$ at the event horizon
is plotted as a function of the coupling $\alpha$ for  $r_h=0.10$, $b=c=m_\pi=1$ and a few values of $a$.
\emph{Right panel:} The value of the Skyrme function $f(r_h)$ at the event horizon (left) is plotted as a function
of the horizon radius $r_h$ for $\alpha=0.05$,  $c=m_\pi=1$ and some set of decreasing
values of $a=b$.}
\end{figure}

Evidently, the critical value of the horizon radius $r_h^{cr}$ is decreasing and approaching zero,
as both parameters $a$ and $b$ tend to zero. Thus, we confirm that hairy black holes
do not exist in the BPS Einstein-Skyrme model.

\section{Summary}

The main result of the paper is the observation that there are Skyrme type models with the usual $\mathbb{S}^3$
target space which do not possess hairy black hole solutions, even though they support solitons (Skyrmions) both in 
flat and dynamically curved (coupled to gravity) space-time. As an example, we considered the BPS Skyrme model with
a single vacuum potential, for which we analytically proved the non-existence of such hairy black hole solutions.

Secondly, we found strong evidence for the conjecture that, within the general Skyrme model, hairy black holes
 exist if the following two conditions are fulfilled:
\begin{enumerate}
\item the generalized Skyrme model contains the Skyrme term i.e., the four derivative term $\mathcal{L}_4$, and
\item the model admits solitonic solutions in flat space-time.
\end{enumerate}
This distinguished role of the Skyrme term for the existence of hairy black hole is  an intriguing fact,
which may have interesting applications if it continues to hold for higher values of the topological charge.
As self-gravitating Skyrmions with sufficiently high baryon number should describe neutron stars, the Skyrme term might then control how much
nuclear matter can escape from the gravitational collapse of a neutron star (or several colliding compact stars). Further, in a recent publication \cite{dvali} it was speculated that the existence of hairy Skyrmion black holes might imply that the conservation of baryon number can survive the gravitational collapse to a black hole. In the light of our results, this leads to the obvious question which role is played by the Skyrme term in this process.

Thirdly, we found strong numerical evidence for the conjecture that, with the sextic term $\mathcal{L}_6$ present (i.e., for $c\not= 0$), the hairy black hole solutions along the unstable (higher mass) branch do not exist all the way down from $r_h = r_h^{\rm cr}$ to $r_h =0$. They only exist for an interval $0< r_h^{\rm ex} \le r_h\le r_h^{\rm cr}$, where for $r_h =r_h^{\rm ex}$ the metric function $\sigma$ develops a zero at the horizon. In physical terms, this corresponds to an extremal (zero temperature) hairy black hole.

Finally, we found a new type of the hairy black holes with compact hair.
These solutions of the $\mathcal{L}_0+\mathcal{L}_4+\mathcal{L}_6$ submodel
have a number of interesting features, whose systematic study is, however, beyond the purpose of this work.

Strictly speaking, our results based on numerical calculations have been demonstrated only for the pion mass potential. We think, however, that they continue to hold for a more general class of potentials (e.g., for one-vacuum potentials, like the exact analytical result for the BPS submodel). In other words, the qualitative behaviour of hairy black hole solutions is, to a certain degree, independent of the potential and determined by the quartic Skyrme term and by the sextic term $\mathcal{L}_6$.

\vspace*{0.2cm}

There are many directions in which the present work can be further continued. One can ask what happens in
higher dimensions. In general, in a $(d+1)$ dimensional space-time one can consider a generalized Skyrme field
$\vec{\phi} \in \mathbb{S}^d$. Then, the question is for which higher dimensional generalizations of the Skyrme model
black holes carry Skyrmionic hair. In other word, which term, among all $(d+1)$ dimensional Lorentz invariant terms
maximally quadratic in time derivatives, is crucial for the existence of hairy black holes. Is it still the four-derivative term which plays an essential role?

Independently, one can ask how the (non)existence of hairy Skyrmionic black holes is influenced by some modifications
of gravity as, for example,  Gauss-Bonnet or dilaton-Gauss-Bonnet \cite{GB} gravity.

One should also verify what happens if, instead of static black holes, one would consider spinning Kerr solutions, and the corresponding spinning Skyrmions (see, e.g., \cite{spinning}).
Does it lead to the appearance of Skyrmionic hair for submodels which do not have such solutions in the static limit, as happens, e.g. in scalar field models (see, e.g., \cite{radu})?

Finally, it might also be interesting to study how our results change in the presence of a non-zero cosmological constant, analogously to what was done, e.g., in \cite{Shiiki:2005aq} for the case of the standard Skyrme model.
\\ \\ 
{\bf Notes added in proof:}

1. Simultaneously with our paper, an independent investigation of the same 
issue (black holes in the general Skyrme model, although for a different 
potential), appeared on the arXiv (now published in \cite{bjarke2}). The results of that paper are 
completely compatible with ours, but the authors of \cite{bjarke2} made an 
observation which equally holds in our case, so we want to comment on it 
briefly. Concretely, they found (for their potential choice) that hairy BH 
solutions along the unstable branch exist all the way down to $r_h =0$ not 
only for $c=0$ but also for sufficiently small, nonzero $c$. Let us 
comment that the same happens in our case, as may be understood easily. Indeed, from our Fig. 2 (left) it follows that, for $\alpha = 0.05$ (the 
value we chose for our BH numerical solutions) a regular Skyrmion solution 
on the upper (unstable) branch exists for $c=0$ but not for $c=0.01$. But 
upon closer inspection of the same figure, it is plausible that the 
regular upper branch solution will exist for sufficiently small but 
nonzero $c$. The numerical calculation shows that this solution exists for 
$0\le c \le c_1 $where $c_1 \sim 0.006$. But this means that for this 
range of values there exists a regular $r_h = 0$ solution which BH solutions on the upper branch  may approach in the limit $r_h \to 0$, and this is precisely what happens, as a detailed numerical study for these 
small values of $c$ reveals. 

2. In Section II we gave an exact proof that the BPS submodel cannot 
support hairy BH solutions, but here we want to mention that there exists 
a more physical argument leading to the same conclusion. The 
energy-momentum tensor of the BPS Skyrme model is the energy-momentum 
tensor of a perfect fluid and, in the static case, may be expressed in 
terms of the energy density $\epsilon$ and pressure $P$. For the axially 
symmetric ansatz (12), the explicit expressions are $\epsilon = \epsilon_6 
+ \epsilon_0$, where
\begin{equation} \nonumber 
\epsilon_6 = \frac{cB^2N}{4\pi^4 r^4}\sin^4 f f'^2 \; , \quad \epsilon_0 
= \mu^2 \mathcal{U} ,
\end{equation} 
and $P =\epsilon_6 - \epsilon_0$. But at the horizon $r_h$, $\epsilon_6 
(r_h)=0$ because the metric function $N(r_h )=0$ by definition. This 
implies that the pressure is {\em negative} at the horizon, which is 
incompatible with the existence of a stable, static configuration.

\section*{Acknowledgements}
The authors acknowledge financial support from the Ministry of Education, Culture, and Sports, Spain (Grant No. FPA 2014-58-293-C2-1-P), the Xunta de Galicia (Grant No. INCITE09.296.035PR and Conselleria de Educacion), the Spanish Consolider-Ingenio 2010 Programme CPAN (CSD2007-00042), and FEDER. 
Y.S. gratefully
acknowledges support from the Russian Foundation for Basic Research
(Grant No. 16-52-12012), DFG (Grant LE 838/12-2) and JINR Bogoljubov-Infeld Programme.
O.K. gratefully acknowledges support
by FP7, Marie Curie Actions, People,
International Research Staff Exchange Scheme (IRSES-606096)
and by the DFG within the Research
Training Group 1620 ``Models of Gravity''.
A.W. thanks Piotr Bizon, Jutta Kunz and Burkhard Kleihaus for discussion. Y.S.
would like to thank Jutta Kunz and Burkhard Kleihaus for discussions and Eugen Radu for useful comments and suggestions on
numerical calculations for this paper.

\end{document}